\pdfoutput=1
\documentclass[11pt,twoside,a4paper]{article}

\usepackage[utf8]{inputenc}
\usepackage[sort&compress,numbers,colon,merge]{natbib}
\usepackage{caption}
\usepackage{subcaption}
\usepackage{tikz}
\usetikzlibrary{mindmap,trees}
\usetikzlibrary{decorations.shapes}

\usepackage{a4wide}
\usepackage{fancyhdr}
\usepackage{graphicx}
\usepackage{url}
\usepackage{amsfonts}
\usepackage{amsmath,amssymb,amsthm}
\numberwithin{equation}{section}

\usepackage{mathrsfs}

\usepackage{multirow}
\usepackage{color}
\usepackage[colorlinks]{hyperref}
\usepackage{cancel}
\usepackage{bbm}
\hypersetup{
pdfauthor={Martin Holthausen, Manfred Lindner and Michael A. Schmidt},
pdftitle={CP and Discrete Flavour Symmetries}
}

\makeatletter
\renewcommand{\fnum@table}{\textbf{\tablename~\thetable}}
\renewcommand{\fnum@figure}{\textbf{\figurename~\thefigure}}
\makeatother

\newcommand{\im}{\ensuremath{\mathrm{Im}}}
\newcommand{\tr}{\ensuremath{\mathrm{tr}}}

\newcommand{\diag}{\ensuremath{\mathrm{diag}}}
\newcommand{\I}{\ensuremath{\mathrm{i}}}

\newcommand{\braket}[1]{\ensuremath{\left<#1\right>}}
\newcommand{\ev}[1]{\ensuremath{\left\langle#1\right\rangle}}
\newcommand{\vev}[1]{\ensuremath{\left\langle#1\right\rangle}}
\newcommand{\hc}{\ensuremath{\text{h.c.}}}

\newcommand{\be}{\begin{equation}}
\newcommand{\ee}{\end{equation}}
\newcommand{\ba}{\begin{eqnarray}}
\newcommand{\ea}{\end{eqnarray}}
\newcommand{\SU}[1]{\ensuremath{\mathrm{SU}(#1)}}

\newcommand{\SO}[1]{\ensuremath{\mathrm{SO}(#1)}}

\newcommand{\conj}[1]{\ensuremath{\mathrm{conj}(#1)}}
\newcommand{\id}{\ensuremath{\mathrm{id}}}
\newcommand{\Eqref}[1]{Eq.~(\ref{#1})}

\newcommand{\Figref}[1]{Fig.~\ref{#1}}
\newcommand{\Tabref}[1]{Tab.~\ref{#1}}
\newcommand{\Secref}[1]{sec.~\ref{#1}}

\newcommand{\Rep}[1]{\ensuremath{\underline{\mbox{\textbf{#1}}}}}
\newcommand{\MoreRep}[2]{\ensuremath{\underline{\mbox{\textbf{#1}}} _{\mbox{\textbf{#2}}}}}
\newcommand{\SG}[2]{\ensuremath{\mathrm{SG}(#1,#2)}}

\newcommand{\GAP}{\texttt{GAP}}

\newcommand{\aut}[1]{\ensuremath{\mathrm{Aut}(#1)}}
\newcommand{\out}[1]{\ensuremath{\mathrm{Out}(#1)}}
\newcommand{\inn}[1]{\ensuremath{\mathrm{Inn}(#1)}}
\newcommand{\zentrum}[1]{\ensuremath{\mathrm{Z}(#1)}}

\newcommand{\abs}[1]{\ensuremath{\left\vert#1\right\vert}}

\begin{document}
\allowdisplaybreaks[1]

\newcounter{nodecount}
\newcommand\tabnode[1]{\addtocounter{nodecount}{1} \tikz \node (\arabic{nodecount}) {#1};}

\tikzstyle{every picture}+=[remember picture,baseline]
\tikzstyle{every node}+=[inner sep=0pt,anchor=base,
text depth=.25ex,outer sep=1.5pt]
\tikzstyle{every path}+=[thick, rounded corners]
\tikzset{
  plabel/.style={inner sep=2pt}
}


\begin{titlepage}

\begin{center}
{\Huge\sffamily\bfseries 
CP and Discrete Flavour Symmetries
}
\\[10mm]
{\large
Martin Holthausen\footnote{\texttt{martin.holthausen@mpi-hd.mpg.de}}$^{(a)}$, Manfred Lindner\footnote{\texttt{lindner@mpi-hd.mpg.de}}$^{(a)}$ and 
Michael A.~Schmidt\footnote{\texttt{michael.schmidt@unimelb.edu.au}}$^{(b)}$}
\\[5mm]
{\small\textit{$^{(a)}$
Max-Planck Institut f\"ur Kernphysik, Saupfercheckweg 1, 69117
Heidelberg, Germany
}}\\
{\small\textit{$^{(b)}$
ARC Centre of Excellence for Particle Physics at the Terascale,
School of Physics, The University of Melbourne, Victoria 3010, Australia
}}

\end{center}
\vspace*{1.0cm}
\date{\today}

\begin{abstract}
\noindent
We give a consistent definition of generalised CP transformations in the context of discrete flavour symmetries. Non-trivial consistency conditions imply that every generalised CP transformation can be interpreted as a representation of an automorphism of the discrete group. This allows us to give consistent generalised CP transformations of popular flavour groups. We are able to clear up issues concerning recent claims about geometrical CP violation in models based on $T^\prime$, clarify the origin of "calculable phases" in $\Delta(27)$ and explain why apparently CP violating scalar potentials of $A_4$ result in a CP conserving ground state.
\end{abstract}
\tableofcontents
\end{titlepage}

\setcounter{footnote}{0}

\section{Introduction}
After the discovery of a sizeable value of $\theta_{13}$ by the reactor experiments DoubleChooz\cite{Abe:2011fz}, DayaBay\cite{An:2012eh} and RENO\cite{Ahn:2012nd} the door has been pushed wide open to measure the last undetermined parameters of the Standard Model, namely the CP phases of the lepton sector. Of special interest is the Dirac CP-phase $\delta_{CP}$ as it can be experimentally determined in neutrino oscillation experiments in the foreseeable future.\footnote{ To discern Majorana phases from possible future signals of neutrinoless double beta decay experiments will always be model dependent and thus seem less promising. }

In the lepton sector, there is the proud/infamous tradition to explain the structure of mixing angles through the introduction of non-abelian discrete symmetries. The relative lack of success with regard to the reactor angle $\theta_{13}$ has not deterred the field from using the same set of ideas to try and predict the missing CP phase $\delta_{CP}$ using discrete symmetries. For example, a geometrical origin of the CP phase has been discussed for the group $\Delta(27)$~\cite{Branco:1983tn,*deMedeirosVarzielas:2011zw,*de-Medeiros-Varzielas:2012fk,*Varzielas:2012pd,*Bhattacharyya:2012pi,Ferreira:2012ri} and there have been attempts to explain CP violation as a result of complex Clebsch-Gordon coefficients in models based on the group $T^\prime$~\cite{Chen:2009lr,Meroni:2012ty}.  
However, sometimes definitions of CP have been used that are incompatible with the discrete flavour symmetry, leading to inconsistencies, as will be discussed in detail later. In order to relate CP violation to the complex Clebsch-Gordan coefficients, a CP symmetry has to be imposed on the Lagrangian, which is then broken spontaneously~\cite{Lee:1973iz,*Branco:1979pv}.\footnote{ Recently, a general group-theoretic condition for spontaneous CP violation has been given in Ref.~\cite{Haber:2012np}.} 

We here give a consistent general definition of CP transformations in the context of non-abelian discrete flavour groups. We will show that in many cases it is not possible to define CP in the naive way, $\phi \rightarrow \phi^*$, but rather a non-trivial transformation in flavour space is needed. Indeed there is a one-to-one correspondence between generalised CP transformations~\cite{Ecker:1981wv,*Ecker:1983hz,*Neufeld:1987wa} and the outer automorphism group of the flavour group. It should not be surprising that outer automorphisms play a role in the definition of CP as complex conjugation is an outer automorphism of the field of complex numbers and the definition of CP transformations as automorphisms in the context of gauge theories has been discussed long ago by Grimus and Rebelo~\cite{Grimus:1997fk}.
Generalised CP transformations in the context of discrete symmetries have been used before in Ref.~\cite{Harrison:2002et,*Harrison:2002kp,*Harrison:2004he,*Grimus:2003yn,*Farzan:2006vj,*Joshipura:2009tg,*Grimus:2012hu,*Mohapatra:2012tb,*Krishnan:2012me,Feruglio:2012cw}.

While the outer automorphism group of continuous groups is either trivial or a $Z_2$ (with the sole exception of $\SO{8}$, whose outer automorphism group is $S_3$), the outer automorphism group of discrete groups can be very rich. For example the well-known flavour group $\Delta(27)$ has an automorphism group of order $432$.

As a result of our investigation of generalised CP transformations, we present consistent definitions of CP for all groups of order smaller than 31 that contain three dimensional representations. 
Highlights are the case of $T^\prime$, where we show that there is one consistent definition of CP, which we apply to the models discussed in Ref.~\cite{Chen:2009lr,Meroni:2012ty}. We show that this CP is spontaneously broken by the VEVs of the doublets and it is additionally explicitly broken by the phases of Yukawa couplings and therefore the results obtained have to be considered as unphysical and basis dependent.
For the group $\Delta(27)$ we are able to explain the so-called calculable phases as a result of an accidental generalised CP symmetry that had so far been overlooked in the literature~\footnote{Accidental CP symmetries have also been observed in scalar potentials in models based on dihedral groups $D_{n}$ and its double cover $Q_n$~\cite{Babu:2004tn,Babu:2011mv}.}.

The outline of the paper is as follows. In \Secref{sec:general}, we define a generalised CP transformation and discuss its connection with the outer automorphism group. The implications of a generalised CP transformation for the physical phases are discussed in \Secref{sec:physicalPhases}. In \Secref{sec:examples}, we apply our general considerations to specific examples. In order to uniquely specify each group, we denote it by $\SG{O}{N}$ with $O$ being its order and $N$, the number in the \GAP~\cite{GAP4} SmallGroups catalogue~\cite{SmallGroups:2011}. In particular, we will discuss all groups of order less than $31$ with a three-dimensional representation. Finally, we conclude in \Secref{sec:conclusions}.

For the convenience of the reader, we will briefly define all relevant group theoretical notions in the text or in a footnote. More detailed knowledge can be gained from standard group theory text books. See \cite{Fairbairn:1982jx,*Grimus:2011ff, *Ishimori:2012zz} for an overview of discrete groups, which have been used in the context of flavour symmetries.


\section{Generalised CP and the Outer Automorphism Group\label{sec:general}}
In order to simplify the discussion, we will focus on finite discrete groups only. We do not consider the transformation under the Lorentz group or any continuous symmetry group and therefore restrict ourselves to scalar multiplets unless stated otherwise. 
An extension to higher spin representations of the Lorentz group and continuous groups is straightforward.
Let us consider a scalar multiplet
\begin{align}
\phi=\left( \begin{array}{ccccc}\varphi_R,& \varphi_P,& \varphi_P^*,& \varphi_C,& \varphi_C^* \end{array}\right)^T
\label{eq:vector-of-reps.}
\end{align}
that contains fields in real(R), pseudo-real(P) and complex(C) representations of the discrete group $G$. Note that $\phi$ always contains the field and its complex conjugate. The discrete group $G$ acts on $\phi$ as 
\begin{align}
\phi\stackrel{G}{\longrightarrow} \rho(g) \phi, \qquad g \in G.
\label{eq:group-trafo}
\end{align}
where $\rho$ is a representation $\rho:G\rightarrow GL(N, \mathbb{C})$, which is generally reducible. In fact $\rho(G)\subset U(N)$, since we are only considering unitary representations. The representation $\rho$ decomposes in a block diagonal form
\begin{equation}
\rho=
\begin{pmatrix}
\rho_R &&&&\\
&\rho_P&&&\\
&&\rho_P^*&&\\
&&&\rho_C&\\
&&&&\rho_C^*\\
\end{pmatrix}\;.
\end{equation}
A {\em generalised CP transformation} has to leave $\vert \partial \phi\vert^2$ invariant and thus is of the form 
\begin{equation}
\phi\stackrel{CP}{\longrightarrow}U\phi^*
\label{eq:def-gen-CP-U}
\end{equation}
with $U$ being a unitary matrix, which is not necessarily block-diagonal, because it generically interchanges representations. Even different real representations can be connected by such a CP transformation, as we will discuss later. 

If the representation is real, i.e. $\rho=\rho^*$, there is always the trivial CP transformation $\phi\to\phi^*$, which acts trivially on the group. In the following, we will take $\rho$ to be complex and \emph{faithful}, i.e.~$\rho$ is injective. If $\rho$ were not faithful then the theory would only be invariant under the smaller symmetry group isomorphic to $G/\ker \rho$ and the restricted representation would be faithful. 

\begin{figure}[tb]\centering
\begin{tikzpicture}[scale=4]
\draw (-1,0) node (ginG) {$g \in G$};
\draw (0,.5) node (rhog) {$\rho(g)^*$};
\draw (1,.5) node (UrhoU){$U \rho(g)^* U^{-1}=\rho(g')$};
\draw (2,0) node (ug){$u(g)=g' \in G$};
\draw[->,out=90,in=180] (ginG) to node[above,yshift=1ex]{$\rho$}  (rhog);
\draw[->,out=0,in=180] (rhog) to  (UrhoU);
\draw[->,out=0,in=90] (UrhoU) to node[right,xshift=2ex,yshift=1ex]{$\rho^{-1}$}  (ug);
\draw[->,out=0,in=180] (ginG) to node[above,yshift=1ex]{$u:G\rightarrow G$}  (ug);
\end{tikzpicture}
\caption{The matrix U that appears in the definition of CP defines an automorphism $u:G\rightarrow G$ of the group G.}
\label{fig:aut-definition}
\end{figure}
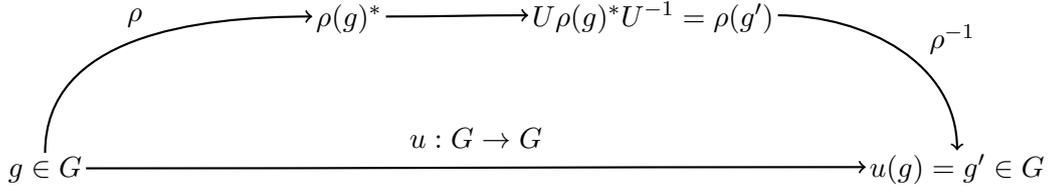

Note that \Eqref{eq:def-gen-CP-U} in combination with \Eqref{eq:vector-of-reps.} implies the existence of a matrix $W$ with $W^2=\mathbbm{1}$, which exchanges the complex conjugate components of $\phi$,
\begin{equation}
\phi^*=W \phi\;,\;\;\mathrm{which\, implies}\;\; 
\rho(g)=W \rho(g)^*W^{-1}\;.
\label{eq:W-def}
\end{equation}
See \Secref{sec:Z3} and especially \Eqref{eq:Z3defW} for a concrete example.
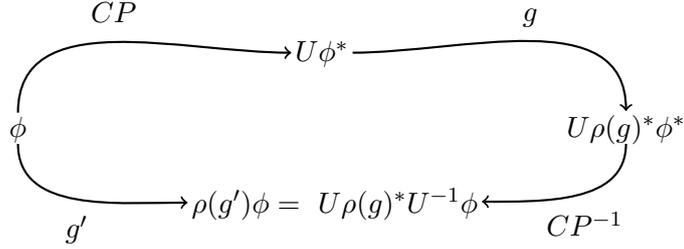
\begin{figure}[bt]\centering
\begin{tikzpicture}[scale=2]
\draw (0,0) node (phi) {$\phi$};
\draw (2*1,.5) node (Uphi) {$U \phi^*$};
\draw (2*2,0) node (Urhophi) {$U \rho(g)^*\phi^*$};
\draw (1.5,-.5) node (rhophi) {$\rho(g') \phi=$};
\draw (2.5,-.5) node (UrhoUphi) {$U \rho(g)^* U^{-1} \phi$};
\draw[->,out=90,in=180] (phi) to node[above,yshift=1ex]{$CP$}  (Uphi);
\draw[->,out=0,in=90] (Uphi) to node[above,yshift=1ex]{$g$}  (Urhophi);
\draw[->,out=-90,in=180] (phi) to node[below,yshift=-1ex]{$g'$} (rhophi) ;
\draw[->,out=-90,in=0] (Urhophi) to node[below,yshift=-1ex,xshift= 3pt]{$CP^{-1}$} (UrhoUphi.east) ;
\end{tikzpicture}
\caption{CP definition.}
\label{fig:CP-definition}
\end{figure}
Comparing first performing a group transformation and then performing a CP transformation with the inverse order of operations and demanding that the resulting transformation is contained in the symmetry group $G$ of the theory, as shown in \Figref{fig:CP-definition}, one finds the requirement that
\begin{align}
U \rho(g)^* U^{-1} \in  \im \rho\equiv\rho(G)\;,
\label{eq:CP-def}
\end{align}
i.e.~the CP transformation maps group elements $\rho(g)$ onto group elements $\rho(g^\prime)$. We will refer to this condition as consistency condition and denote models satisfying this condition consistent. If the condition \eqref{eq:CP-def} is not fulfilled, the group $G$ is not the full symmetry group of the Lagrangian and one would have to consider the larger group, which closes under CP transformations \eqref{eq:CP-def}. We do not consider this case further and will assume that the group $G$ is the full symmetry group of the Lagrangian.
Hence, a generalised CP transformation preserves the group multiplication, i.e.~ $U\rho(g_1 g_2 )^* U^{-1}=U\rho(g_1)^* U^{-1} U\rho(g_2 )^* U^{-1}$ and $U\mathbbm{1}^*U^{-1}=\mathbbm{1}$, and therefore is a homomorphism~\footnote{A \emph{ (group) homomorphism} $\mu : G\rightarrow H$ is a mapping preserving the group structure, i.e.~$\mu(g_1 g_2)=\mu(g_1)\mu(g_2)\;\forall g_{1,2}\in G$, $\mu(g^{-1})=\mu(g)^{-1}$, and $\mu(E_G)=E_H$, where $E_{G,H}$ denotes the identity elements of $G$ and $H$, respectively.}. Furthermore the CP transformation is bijective, since $U$ is unitary and therefore invertible. Hence, CP is an automorphism~\footnote{An \emph{automorphism} $\mu$ of a group $G$ is a bijective homomorphism $\mu : G\rightarrow G$.} of the group, as is depicted in \Figref{fig:aut-definition}. 

Indeed, \emph{the possible matrices $U$ of Eq. \eqref{eq:CP-def} form a representation of the automorphism group~\footnote{The \emph{automorphism group} $\aut{G}$ is the set of all automorphisms of $G$ with composition as group multiplication.} $\aut{G}$ of $G$}, which we are showing in the following. 

$U$ represents the automorphism $u:G\rightarrow G$ given by
\begin{equation}
u: g \in G \rightarrow \rho(g)\rightarrow U \rho(g)^* U^{-1}=\rho(g^\prime)\rightarrow g^\prime=\rho^{-1}(U \rho(g)^* U^{-1}) \in G 
\end{equation}
or
\begin{align}
U \rho(g)^*U^{-1}=\rho(u(g))\;.
\label{eq:aut-def-from-U}
\end{align}
It is straightforward to show that this mapping $u:G\to G$ is an automorphism, indeed.

Vice versa, if $u:G\to G$ is an automorphism, we can explicitly construct a matrix $U$ in the following way.
We first extend $G$ to a group $G^\prime$ containing $G$ as a normal subgroup and $u(g)=g' g g'^{-1}\; \forall g \in G$ with $g' \in G^\prime$. This can be achieved as follows.
Taking the order of $u$~\footnote{The \emph{order} of a group element $u$ of $G$ is given by the smallest $n\in \mathbbm{N}$ with $u^n=\mathrm{id}_G$.} to be $\mathrm{ord}(u)=n$, we define the homomorphism 
\begin{equation}
\theta: Z_n=(\{0,..,n-1\},+)\rightarrow \aut{G}: 1\to \theta_1\equiv u\;,
\end{equation}
which has a trivial kernel.
This homomorphism thus defines the semi-direct product group $G'=G\rtimes_\theta Z_n$ with the group multiplication
\begin{equation}
(g_1,z_1)\star  (g_2,z_2)=(g_1\theta_{z_1}(g_2),z_1+ z_2 )\;.
\end{equation}
Keeping track of the multiplication rules, we find
\begin{equation}\label{eq:CPinSemidirectProd}
(E,1)\star (g,z)\star(E,1)^{-1}=(u(g),z)\;,
\end{equation}
where $E$ is the identity element of $G$. 
The outer~\footnote{An \emph{inner automorphism} $\mu_h$ of a group $G$ is an automorphism, which is represented by conjugation with an element $h\in G$, i.e. $\mu_h\equiv\conj{h}: g\to hgh^{-1}$. If an automorphism can not be represented by conjugation with a group element, it is called an \emph{outer automorphism}.} automorphism $u$ of $G$ becomes an inner automorphism of $G'$ and we can obtain a matrix representation of $u$ (or equivalently $(E,1)$) by the standard techniques for finding matrix representations of groups, for example by using the computer algebra system \GAP~\cite{GAP4}. In order to relate the matrix representation of $u$ in the semidirect product group $G^\prime$ with the matrix $U$ in the CP transformation of a representation $\rho$ of $G$, we have to consider a representation $\rho^\prime$ of $G^\prime$ whose restriction to the subgroup $G$ is the representation $\rho$\footnote{There is always a representation $\rho^\prime$, whose restriction to $G$, $\left.\rho^\prime\right|_G$, contains $\rho$. If $\left.\rho^\prime\right|_G\neq \rho$, one has to extend $\rho$ to $\left.\rho^\prime\right|_G$, because the CP transformation connects the representation $\rho$ to another representation.}, i.e. $\left.\rho^\prime\right|_G= \rho$. 
In terms of the representation $\rho^\prime$, \Eqref{eq:CPinSemidirectProd} translates to
\begin{equation}
\rho^\prime((E,1)) \rho^\prime((g,z)) \rho^\prime((E,1)^{-1}) = \rho^\prime((u(g),z))\;,
\end{equation}
which can be written in terms of a CP transformation
\begin{equation}\label{eq:CPGprime}
U \rho^{\prime}((g,z))^* U^{-1} = \rho^\prime((u(g),z))
\end{equation}
with $U=\rho^\prime((E,1)) W$ using the matrix $W$ introduced in \Eqref{eq:W-def}. Finally, we have to consider the restriction of $\rho^\prime$ to the subgroup $G$, i.e. $\left.\rho^\prime\right|_G = \rho$ and therefore
\begin{equation}
U \rho(g)^* U^{ -1} = \rho(u(g))\;.
\label{eq:cons-cond}
\end{equation} 
with $\rho(g)=\rho^\prime((g,0))$. Hence there is a one-to-one correspondence between set of matrices $U$ and the automorphism group of $G$. If $\rho$ does not contain all representations that are connected via the outer automorphism $u$, there is no matrix $U$ that fulfils the consistency condition (\ref{eq:cons-cond}). To implement this CP transformation, the vector $\phi$ therefore has to be enlarged by the missing representations.
For example in case of the group $Q_8\rtimes A_4$, which we discuss in \Secref{sec:CP-Q8}, different real representations are interchanged by the matrix $U$ and the CP transformation can only be implemented if all representations connected by the corresponding automorphism are present in the theory. The same is true for the group $A_5$ where the two different real 3-dimensional representations are interchanged by the outer automorphism of the group.

The automorphisms form a group with composition as group multiplication, i.e.~$u^\prime=\tilde{u}\circ u$ is again an automorphism represented by 
\begin{equation}
U^\prime \rho(g)^*{U^\prime}^{-1}=\rho({u^\prime}(g))
\end{equation}
with 
\begin{equation}
\rho(u^\prime(g))=\rho(\tilde{u}(u(g)))=\tilde{U}W\rho(u(g))W\tilde{U}^{-1}=\tilde{U}WU\rho(g)^*{U^{-1}}W\tilde{U}^{-1}
\end{equation}
and thus 
\begin{align}
U^\prime=\tilde{U}WU.
\label{eq:group-structure-aut}
\end{align}
The trivial automorphism $\id(g)=g \;\forall g \in G$ is represented by $U=W$ and the inverse automorphism $u^{-1}$ is represented by
$W U^{-1}W$. We thus have a homomorphism from the automorphism group to the group of matrices $U$ defined in \Eqref{eq:def-gen-CP-U} with the conjunction $\star$: $(A,B)\to A\star B\equiv A W B$. With respect to this conjunction the matrices $U$ form a representation of the automorphism group.\hfill$\blacksquare$

For any solution $U$ of \Eqref{eq:CP-def} the matrix $\rho(g)U$ is also a solution for any $g \in G$, which corresponds to performing a CP transformation followed by a group transformation described by $\rho(g)$. The group transformation corresponds to an inner homomorphism, which does not pose any new restrictions~\footnote{There are interesting phenomenological consequences for inner automorphisms, if the CP symmetry is left unbroken in one sector of the theory like the neutrino sector as discussed in~\cite{Feruglio:2012cw}. However, we are more interested in the consistent definition in the unbroken theory and therefore do not further consider inner automorphisms.}. It is therefore sufficient to consider automorphisms with inner automorphisms modded out. \emph{Hence the group of generalised CP transformations is given by the outer automorphism group, which is defined by}
\begin{equation}
\out{G}\equiv \aut{G}/\inn{G}\;,
\end{equation}
where $\inn{G}$ denotes the \emph{inner automorphism group}~\footnote{For every group G there is a natural group homomorphism $G\rightarrow\aut{G}$ whose image is $\inn{G}$ and whose kernel is the \emph{centre of G}, $\zentrum{G}$, i.e. the subset of $G$ which commutes with all elements of $G$. In short 
$
\inn{G}\cong G/\zentrum{G}\;.
$
Thus, if $G$ has trivial centre it can be embedded into its own automorphism group.}, the set of all inner automorphisms.
Moreover, since the invariance under a CP transformation leads to the invariance under the subgroup generated by CP, the physically distinct classes of CP are given by the subgroups of the outer automorphism group.

As we will be using the character table in the discussion of the different groups (See e.g.~\Tabref{tab:chTabZ3}), we will briefly comment on how automorphisms act on the character table. As automorphisms are mappings from the group into itself and there is a unique character table for each group up to reordering of rows and columns, automorphisms are symmetries of the character table and can not change the character table besides exchanging rows and columns. While the action of the automorphisms on conjugacy classes should be self-explanatory, the action on representations requires further comment: Note that \Eqref{eq:aut-def-from-U} may be read as a similarity transformation between the representations $\rho^*$ and $\rho\circ u$. By composition an automorphism therefore interchanges representations. This is the reason why we have chosen to define CP on the reducible representation shown in \Eqref{eq:vector-of-reps.}. 

Inner automorphisms act via conjugation on the group. Hence, they map elements of the individual conjugacy class onto elements of the same conjugacy class. Neither do they exchange representations and therefore they do not change the character table. Outer automorphisms on the other hand map elements from one conjugacy class to another as well as one representation to another and thus exchange rows and columns. 

Finally, note that it is always possible to use the freedom of a multiplication with an arbitrary phase factor. Hence if $\phi\to U\phi^*$ is a generalised CP transformation, so is
\begin{equation}
\phi\rightarrow e^{i\alpha}U\phi^*\;.
\end{equation}
This does not lead to any additional constraints but only changes the phase factors of the different couplings.

We follow \cite{Grimus:1997fk} and call a basis where $U$ may be represented by the identity matrix times a phase, $\phi\rightarrow e^{\I \alpha}\phi^*$, a \emph{CP basis}. Note that under a change of basis $\phi^\prime=V \phi $ we have
\begin{align}
\phi^\prime\rightarrow (V U V^T)\phi^\prime
\label{eq:CP-basis}
\end{align}
and it is thus not always possible to perform a basis change to a CP basis where $V U V^T$ is diagonal~\cite{Branco:2012hc}, but it is possible to go to a basis where $U$ takes the form \cite{Ecker:1987qp}
\begin{equation}
V U V^T=\begin{pmatrix}O_1&&&\\&\ddots&&\\&&O_l&\\&&&\mathbbm{1}_m\end{pmatrix}
\end{equation}
with $2l+m=\mathrm{dim}(U)$ and $O_i$ being orthogonal $2\times2$ matrices.

\section{Physical Implications of a Generalised CP Symmetry\label{sec:physicalPhases}}

The existence of a generalised CP symmetry implies that there is no direct CP violation and CP violation can only be generated via spontaneous symmetry breaking. This has been studied in terms of weak basis invariants~\cite{Gronau:1986xb,*Bernabeu:1986fc,Branco:1986gr}. 
A necessary and sufficient set of weak basis invariants, which measure the CP violation in the lepton sector and vanish in the CP conserving case has been proposed in~\cite{Branco:1986gr}. In the following, we will explicitly demonstrate that the weak basis invariant for Dirac CP violation vanishes for our generalised CP symmetry and refer the reader to~\cite{Branco:1986gr} for the remaining weak basis invariants.
Let us consider a left-handed lepton doublet $L=(\nu,e)^T$ with the following mass terms
\begin{equation}
\mathcal{L}_{mass}=-e^T M_e e^c-\frac{1}{2}\nu^T M_\nu \nu+\hc\;.
\end{equation}
It was shown in Ref. \cite{Gronau:1986xb,Bernabeu:1986fc,Branco:1986gr} that Dirac-type CP violation ($\sin\delta_{CP} \neq 0 $) is equivalent to 
\begin{equation}
0\neq \tr \left [H_\nu, H_e \right]^3
\qquad\mathrm{with}\qquad
H_\nu=(M_\nu^\dagger M_\nu)^*\;\;\mathrm{and}\;\; H_e= (M_e M_e^\dagger)^T\;.  
\end{equation}
If $L$ transforms under a generalised CP transformation as 
\begin{equation}
L\stackrel{\mathrm{CP}}{\longrightarrow} U L^C\equiv U\; (\I\sigma_2 L^*)
\end{equation}
where $L^C$ denotes charge conjugation with respect to the Lorentz group and $U$ is unitary, the weak basis invariants $H_{\nu,e}$ have to fulfil 
\begin{equation}
H_\nu=U^T H_\nu^T U^* \qquad H_e= U^T H_e^T U^*
\end{equation}
and therefore (note $[A,B]^T=-[A^T,B^T]$)
\begin{equation}
\tr \left [H_\nu, H_e \right]^3=\tr \; U^T\left [H_\nu^T, H_e^T \right]^3U^*=-\tr  \left[H_\nu, H_e \right]^3=0
\end{equation}
and there is thus no Dirac-type CP violation.


\section{Application to Questions in the Literature\label{sec:examples}}
In the following, we will apply our general discussion to specific groups. We will concentrate on the most popular groups, which have been used in the literature.
\tikzstyle{every picture}+=[remember picture,baseline]
\tikzstyle{every node}+=[inner sep=0pt,anchor=base,
text depth=.25ex,outer sep=1.5pt]
\tikzstyle{every path}+=[thick, rounded corners]

\mathversion{bold}
\subsection{$Z_3\cong\SG{3}{1}$}\label{sec:Z3}
\mathversion{normal}
Let us start the discussion of examples by taking the cyclic group with three elements $Z_3\cong\braket{T \vert T^3=E}$, which is the smallest group with complex representations. There is one non-trivial automorphism $u:T\to T^2$, which is outer and since all group elements commute, there is only the trivial inner automorphism, $\conj{T}=\conj{T^2}=\id.$ The structure of automorphism group is thus:
\begin{align}
\zentrum{Z_3}&\cong Z_3&
\aut{Z_3}&\cong Z_2\\\nonumber
\inn{Z_3}&\cong Z_1&
\out{Z_3}&\cong Z_2
\end{align}
Looking at the character table in \Tabref{tab:chTabZ3},
\begin{table}[tb]\centering
\tikzstyle{every picture}+=[remember picture,baseline]
\tikzstyle{every node}+=[inner sep=0pt,anchor=base,
text depth=.25ex,outer sep=1.5pt]
\tikzstyle{every path}+=[thick, rounded corners]

\setcounter{nodecount}{0}
\begin{tabular}{c|ccc}
& $E$ & \tabnode{$T$} & \tabnode{$T^2$}\\\hline 
\MoreRep{1}{1} & 1 &1 &1\\
\tabnode{\MoreRep{1}{2}} & 1 & $\omega$ & $\omega^2$ \\
\tabnode{\MoreRep{1}{3}} & 1 & $\omega^2$ & $\omega$
\end{tabular}
\begin{tikzpicture}[overlay]
\node [above=.3cm,minimum width=0pt] at (1) (An){};
\node [above=.3cm,minimum width=0pt] at (2) (Bn){};
\draw [<->,out=45,in=135,blue!50, very thick,below=1cm] (An) to (Bn);
\draw [<->,out=225,in=135,blue!50, very thick] (3.west) to (4.west);
\end{tikzpicture}
\caption{Character table of $Z_3$ with $\omega=e^{2\pi\I/3}$. The outer automorphism $u:T\to T^2$ is indicated in blue.\label{tab:chTabZ3}}
\end{table}
we see that the outer automorphism $u:T\to T^2$ indicated in blue acts on the character table by interchanging the conjugacy classes represented by $T$ and $u(T)=T^2$ and the representations $\MoreRep{1}{2}\leftrightarrow \MoreRep{1}{2}\circ u=\MoreRep{1}{3}$, i.e. the rows and columns of the character table, such that the table stays invariant, as an outer automorphism should do. 

Let us consider a theory that contains the complex representation $\varphi\sim \MoreRep{1}{2}$. The vector 
$
\phi=\left(\varphi, \varphi^*\right)^T
$
is acted upon by the group generator $T$ as
\begin{equation}
\rho(T)=\left(\begin{array}{cc} \omega &0 \\ 0&\omega^2 \end{array}\right)
\end{equation}
and we have $\rho(T)^*=\rho(T^2)=\rho(u(T))\in \im \rho$ and therefore  $U=\mathbbm{1}_2$ is a representation of the outer automorphism $u:T\to T^2$. The generalised CP transformation \eqref{eq:def-gen-CP-U} is therefore just the usual $\varphi\rightarrow \varphi^*$.

The matrix $W$ relating the representation $\MoreRep{1}{2}\oplus\MoreRep{1}{3}$ with its complex conjugate is given by
\begin{equation}
\label{eq:Z3defW}
W=\left(\begin{array}{cc} 0&1 \\ 1&0 \end{array}\right)
\end{equation}
and $U=W$ represents the trivial automorphism, or $\varphi \rightarrow \varphi$.

While here it is trivial to find a matrix $U$ representing the outer automorphism $u$, it is instructive to demonstrate the general method of constructing the semi-direct product group $G'=Z_3\rtimes \mathop{CP}$ introduced in \Secref{sec:general} explicitly . It is given by $Z_3\rtimes Z_2$, where $Z_3$ is generated by $T$ and $Z_2$ by the automorphism $u$. Hence, its elements are 
\begin{equation}
\left\{(E,\id),\,
(T,\id),\,
(T^2,\id),\,
(E,u),\,
(T,u),\,
(T^2,u)\right\}
\end{equation}
and the multiplication is defined by 
\begin{equation}
(g_1,u_1)\star (g_2,u_2)\equiv (g_1 u_1(g_2), u_1\circ u_2)\;,
\end{equation}
which defines a non-abelian group of order $6$ and it is isomorphic to $S_3$ being the only non-abelian group of order $6$. It has two generators: $(T,\id)$ and $(E,u)$. The outer automorphism $u: T\to T^2$ of $Z_3$ corresponds to the inner automorphism $\conj{(E,u)}: Z_3\rtimes Z_2\ni g\to (E,u)\star g \star (E,u)^{-1}$ of $Z_3\rtimes Z_2\cong S_3$. The group $S_3$ has three representations: $\MoreRep{1}{1,2}$ and $\Rep{2}$; only the 2-dimensional representation is faithful and the generators are given by
\begin{equation}
\rho_{\Rep{2}}((T,\id))=\left(\begin{array}{cc} \omega &0 \\ 0&\omega^2 \end{array}\right)\;,\qquad\mathrm{and}\qquad 
\rho_{\Rep{2}}((E,u))=\begin{pmatrix}
0&1\\
1&0\\
\end{pmatrix}\;.
\end{equation}
In terms of the subgroup $Z_3=\ev{T}$, it decomposes in the direct sum of the representations $\MoreRep{1}{2}$ and $\MoreRep{1}{3}$ of $Z_3$ with the group generator $\rho(T)=\rho_{\Rep{2}}((T,\id))$.
The automorphism $u$ is represented by the matrix $U^\prime=\rho_{\Rep{2}}((E,u))$ and $\rho(g)\to \rho(u(g))=U^\prime \rho(g) {U^\prime}^{-1}$
and therefore the non-trivial CP transformation belonging to the automorphism $u$ is given by $\rho(g)\to \rho(u(g))=U\rho(g)^*U^{-1}$ with $U=U^\prime W=\mathbbm{1}_2$, as we have found above.
Clearly the trivial automorphism corresponds to $(E, \id)$ and is represented by $U^\prime=\mathbbm{1}_2$ or $U=W$.



\mathversion{bold}
\subsection{$A_4\cong (Z_2\times Z_2)\rtimes Z_3\cong\SG{12}{3}$}
\mathversion{normal}
There is a complete classification of automorphism groups for the alternating groups $A_n$, which is shown in \Tabref{tab:An}. Most of them have a very similar structure. We will discuss the specific case of $A_4=\braket{S,T\vert S^2=T^3=(ST)^3=E}$~\footnote{$A_4$ has been introduced as flavour symmetry in the lepton sector in~\cite{Ma:2001lr}.} in detail. 
\begin{table}[tbh]\centering
\begin{subtable}{0.65\textwidth}\centering
\begin{tabular}{l|cccc}
& $\zentrum{A_n}$ & $\aut{A_n}$ & $\inn{A_n}$ & $\out{A_n}$ \\\hline
$n\geq 4, n\neq6$ & $Z_1$ & $S_n$ & $A_n$ & $Z_2$\\
$n=1,2$ & $Z_n$ & $Z_1$ & $Z_1$ & $Z_1$\\
$n=3$ & $Z_3$ & $Z_2$ & $Z_1$ & $Z_2$\\
$n=6$ & $Z_1$ & $S_6\rtimes Z_2$ & $A_6$ & $Z_2\times Z_2$\\
\end{tabular}
\caption{Structure of the automorphism group of $A_n$\label{tab:An}}
\end{subtable}
\begin{subtable}{.3\textwidth}\centering
\begin{tabular}{l|cccc}
& $E$ & $T$ & $T^2$ & $S$\\\hline 
\MoreRep{1}{1} & 1 &1 &1 &1\\
\MoreRep{1}{2} & 1 & $\omega$ & $\omega^2$ & 1\\
\MoreRep{1}{3} & 1 & $\omega^2$ & $\omega$ & 1\\
\Rep{3} & 3 & 0 & 0 & -1\\
\end{tabular}
\caption{Character Table of $A_4$.\label{tab:A4representations}}
\end{subtable}
\caption{Relevant group structure of the alternating groups $A_n$.}
\end{table}
It is very important for model building and serves as our first non-trivial example. As it can be seen in \Tabref{tab:An}, only the identity element commutes with all other elements and the natural homomorphism $n :A_4\rightarrow \aut{A_4}$ defined by $n(g)=\conj{g}$ is therefore injective. There is one non-trivial outer automorphism $u: (S,T)\to(S,T^2)$. Here and in the following, we only give the action of automorphisms on the generators of the group, which uniquely defines an automorphism.
The character table of $A_4$ is given in \Tabref{tab:A4representations} and it is easy to verify that the automorphism $u$ represents a symmetry of the character table, again interchanging the representations $\MoreRep{1}{2}$ and $\MoreRep{1}{3}$. Let us first discuss the case where we have only one real scalar field in the real representation $\phi\sim\MoreRep{3}{1}$ using the Ma-Rajasekaran\cite{Ma:2001lr} basis:

\begin{align}
\rho_{\MoreRep{3}{1}}(S)=S_3&\equiv\left(\begin{array}{ccc}
1&0&0\\
0&-1&0\\
0&0&-1
\end{array}\right),
& 
\rho_{\MoreRep{3}{1}}(T)=T_3&\equiv\left(\begin{array}{ccc}
0&1&0\\
0&0&1\\
1&0&0
\end{array}\right).
& 
\label{eq:Ma-basis}
\end{align}
In this basis both group generators are real ($\rho(g)^*=\rho(g)\in \im \rho $) and one might be tempted to take $U=\mathbbm{1}_3$ as this fulfils \Eqref{eq:CP-def}.  However, the map derived from $U=\mathbbm{1}_3$ via \Eqref{eq:aut-def-from-U} is  not equal to $u: (S,T)\to(S,T^2)$, but the trivial automorphism $\id_{A_4}$, which is obviously not outer and therefore does not lead to additional constraints on the couplings\footnote{Obviously it still acts non-trivially on the space-time symmetry group as well as possibly the gauge group.}.
One also encounters this problem as soon as one considers contractions such as 
 \begin{equation}
\left( \phi \phi\right)_{\MoreRep{1}{2}}=\frac{1}{\sqrt{3}}\left(\phi_1 \phi_1+\omega^2 \phi_2 \phi_2+\omega \phi_3 \phi_3\right)
\end{equation}
which transform under this "CP" $\phi\rightarrow U \phi^*=\phi$ as 
\begin{equation}
\left( \phi \phi\right)_{\MoreRep{1}{2}}\rightarrow \left( \phi \phi\right)_{\MoreRep{1}{2}}\sim\MoreRep{1}{2}
\end{equation}
which is in conflict with the expectation that CP should involve complex conjugation such that 
\begin{equation}
\left( \phi \phi\right)_{\MoreRep{1}{2}}\rightarrow [\left( \phi \phi\right)_{\MoreRep{1}{2}}]^*\sim\MoreRep{1}{3}.
\end{equation}
Just imagine that the theory contains a real scalar triplet $\chi\sim\Rep{3}$ and a singlet $\xi\sim \MoreRep{1}{3}$. If one defines CP as  $\chi\rightarrow \chi$ and $\xi\rightarrow \xi^*$ then the invariant $\left( \chi \chi\right)_{\MoreRep{1}{2}} \xi$ under CP is mapped to $\left( \chi \chi\right)_{\MoreRep{1}{2}} \xi^*$, which is not invariant under the group and it is forbidden by the combination of $A_4$ and this definition of CP. Looking at this definition of CP, i.e. $\chi\to\chi^*$ and $\xi\to\xi^*$, we can easily check that it does not fulfil the consistency condition in \Eqref{eq:CP-def} and therefore the true symmetry group of the Lagrangian is not $A_4$, but the group generated by $A_4$ and this CP transformation. However, it has been (implicitly) used in a number of works~\cite{Adelhart-Toorop:2010fkv3,Ferreira:2011lr,Machado:2011yq}\footnote{The discussion of CP in Ref.~\cite{Adelhart-Toorop:2010fkv3} has been corrected in Ref.~\cite{Adelhart-Toorop:2010fkv4}.} without properly taking into account the enlarged symmetry group with its additional restrictions on the Lagrangian.

If we instead use the non-trivial solution of \Eqref{eq:CP-def}, which has been discussed in~\cite{Harrison:2002et,*Harrison:2002kp,*Harrison:2004he,*Grimus:2003yn,*Farzan:2006vj,*Joshipura:2009tg,*Grimus:2012hu,*Mohapatra:2012tb,*Krishnan:2012me}
\begin{equation}\label{eq:defU3}
U=U_3\equiv \left(\begin{array}{ccc}
1&0&0\\
0&0&1\\
0&1&0
\end{array}\right)
\end{equation}
that corresponds to the outer automorphism $u:(S,T)\to (S,T^2)$ we immediately see that 
\begin{equation}
\left( \phi \phi\right)_{\MoreRep{1}{2}}\rightarrow [\left( \phi \phi\right)_{\MoreRep{1}{2}}]^*\sim\MoreRep{1}{3}.
\end{equation}
Note that this is the only non-trivial definition of CP (up to inner automorphisms) in any theory that involves the complex representations, since the outer automorphism group is $Z_2$. 

Using \Eqref{eq:CP-def}, we can immediately see that the solution $U=\mathbbm{1}_3$ for $\rho\sim\Rep{3}$ leads to the trivial automorphism $\id_{A_4}$ (up to inner automorphism), when it is extended to the other representations.
Let us consider the vector
$ \phi=( \xi,\xi^*,\chi)^T$ with $\xi\sim \MoreRep{1}{3}$ and  $\chi\sim \MoreRep{3}{1}$ which transforms as  
\begin{align}
\rho(S)=\diag(1,1,S_3)\qquad \rho(T)=\diag(\omega, \omega^2, T_3)
\end{align}
and clearly fulfils $\rho(S)^*=\rho(S)\in \im \rho$ and $\rho(T)^*\notin \im \rho$. We are therefore forced to use $U=\diag(1,1,U_3)$, which gives $U\rho(T)^* U^{-1}=\rho(T^2)\in \im \rho$ and $U\rho(S)^* U^{-1}=\rho(S)\in \im \rho$ and represents the outer automorphism $u:(S,T)\to(S,T^2)$. 
The only consistent (meaning satisfying condition \eqref{eq:CP-def}) non-trivial CP transformation in this theory is thus $\xi \rightarrow \xi^*$ and $\chi \rightarrow U_3 \chi^*=U_3 \chi$. Adding the generator $U$ to $A_4$ results in $S_4$ because $A_4$ can be embedded in $\aut{G}$.

Summarising our discussion, there is only one non-trivial CP transformation (up to inner automorphisms) acting on the reducible representation $\phi\sim\MoreRep{1}{1}\oplus\MoreRep{1}{2}\oplus\MoreRep{1}{3}\oplus\Rep{3}$, which takes the form $\phi\to U\phi^*$ with 
\begin{equation}
U=\begin{pmatrix}
1&0&0&0\\
0&1&0&0\\
0&0&1&0\\
0&0&0&U_3\\
\end{pmatrix}\;.
\end{equation}
The trivial CP transformation corresponding to the trivial automorphism $\id_{A_4}$ is determined by $\phi\to U\phi^*$ with 
\begin{equation}
U=\begin{pmatrix}
1&0&0&0\\
0&0&1&0\\
0&1&0&0\\
0&0&0&\mathbbm{1}_3\\
\end{pmatrix}\;,
\end{equation}
which is equivalent to the transformation $\phi\to\phi$ as can be easily checked.
There are no other CP transformations (up to inner automorphisms).

Since this case is of some relevance to model building, let us dwell on it a bit more and repeat the discussion for the basis 
\begin{equation}
S=\frac{1}{3}\left(\begin{array}{ccc}
-1&2&2\\
2&-1&2\\
2&2&-1
\end{array}\right),\qquad T=\left(\begin{array}{ccc}
1&0&0\\
0&\omega^2&0\\
0&0&\omega
\end{array}\right)
\end{equation}
first used by Altarelli and Feruglio\cite{Altarelli:2006qy}. Here the group elements are complex but the Clebsch-Gordon coefficients are real. The unique result of \Eqref{eq:CP-def} is $U=\mathbbm{1}_3$ up to inner automorphisms. This basis is therefore a CP basis, as defined in Eq. (\ref{eq:CP-basis}). Note that in this basis 
\begin{equation}
\left( \phi \phi\right)_{\MoreRep{1}{2}} = (\phi_2\phi_2 + \phi_1\phi_3 + \phi_3\phi_1),\qquad \left( \phi \phi\right)_{\MoreRep{1}{3}}  = (\phi_3\phi_3 + \phi_1\phi_2 + \phi_2\phi_1)
\end{equation}
and thus
\begin{equation}
\left( \phi \phi\right)_{\MoreRep{1}{2}}\rightarrow [\left( \phi \phi\right)_{\MoreRep{1}{2}}]^*\sim\MoreRep{1}{3}.
\end{equation}
as it should be.

Let us look at a physical situation where a certain confusion about the definition of CP can be alleviated by our definition\footnote{For a related discussion, see \cite{Adelhart-Toorop:2012PhD,Ivanov:2012fp,Degee:2012sk}.}. If one considers the potential for one electroweak Higgs doublet transforming as \MoreRep{3}{1} denoted by $\chi=(\chi_1,\chi_2,\chi_3)^T$ in the basis (\ref{eq:Ma-basis}), there is one potentially complex coupling in the potential \cite{Ma:2001lr,Adelhart-Toorop:2010fkv3,Machado:2011yq}
\begin{align}
\lambda_{5}\, \,  (\chi^\dagger \chi)_{\MoreRep{3}{1}}\left(\chi^\dagger \chi\right)_{\MoreRep{3}{1}}+\hc =\lambda_{5}\, \, \left[ \left(\chi_1^\dagger \chi_2\right)^2+\left(\chi_2^\dagger \chi_3\right)^2+\left(\chi_3^\dagger \chi_1\right)^2\right]+\hc.
\end{align}
It can be easily checked that the generalised CP transformation 
$
\chi\rightarrow U_3 \chi^*
$
acts as 
\begin{equation}
I\equiv\left[ \left(\chi_1^\dagger \chi_2\right)^2+\left(\chi_2^\dagger \chi_3\right)^2+\left(\chi_3^\dagger \chi_1\right)^2\right]\rightarrow \left[ \left(\chi_1^\dagger \chi_2\right)^2+\left(\chi_2^\dagger \chi_3\right)^2+\left(\chi_3^\dagger \chi_1\right)^2\right]=I
\end{equation}
and thus does not give a restriction on the phase of $\lambda_{5}$. Note that the naive CP transformation
$
\chi\rightarrow \chi^*
$
transforms the group invariant $I$ into $I^*$ and therefore restricts $\lambda_{5}$ to be real as was e.g. done in Ref.~\cite{Ferreira:2011lr}. However, we have seen that this naive CP transformation cannot be consistently implemented on the Lagrangian level if there are complex representations, unless it is either the trivial generalised CP transformation, $\id_{A_4}$, or the symmetry group $A_4$ is extended such that it is closed under this naive CP transformation. Therefore it is inappropriate to call the phase of $\lambda_5$ a CP phase. This also explains an observation made in Ref.~\cite{Adelhart-Toorop:2010fkv3}, where it was shown that even for $\arg \lambda_5\neq 0$ the VEV configuration 
\begin{align}
\vev{\chi}=V(1,1,1), \qquad \qquad\vev{\chi}=V(1,0,0)\qquad V \in \mathbbm{R},
\label{eq:CP-VEV}
\end{align}
which of course respects both, the trivial as well as the non-trivial, generalised CP transformations, can be obtained without fine-tuning. This would have been somewhat surprising, as usually symmetry conserving solutions cannot be obtained from explicitly symmetry breaking potentials. However, the phase of $\lambda_5$ does not break the consistent definition of generalised CP, i.e.~does not violate condition \eqref{eq:CP-def}, as does the VEV configuration (\ref{eq:CP-VEV}), therefore everything is consistent.

\mathversion{bold}
\subsection{$T'\cong\SG{24}{3}$}
\mathversion{normal}

The group $T'=\braket{S,T\vert S^4=T^3=(ST)^3=E}\cong \mathrm{SL}(2,3)$~\footnote{$T'$ has been first discussed in a particle physics context in~\cite{Frampton:1994rk}.}, is also an important group in the context of CP violation~\cite{Chen:2009lr,Meroni:2012ty}. It has two elements $\zentrum{T'}=\{ E,S^2\}\cong Z_2$ that commute with all group elements and therefore $\inn{T'}\cong T'/\zentrum{T'}\cong A_4$. There is one non-trivial outer automorphism (up to inner automorphisms) $u:(S,T)\to(S^3,T^2)$. Therefore the automorphism structure can be summarised as:
\begin{align}
\zentrum{T'}&\cong Z_2&
\aut{T'}&\cong S_4\\\nonumber
\inn{T'}&\cong A_4&
\out{T'}&\cong Z_2
\end{align}
A non-trivial CP transformation therefore has to be a representation of $u$ in the sense of \Eqref{eq:CP-def}. Let us now see how it is represented for the various representations of $T^\prime.$

There is a faithful pseudo-real representation 
\begin{equation}
\MoreRep{2}{1}: S = A_1,\qquad T = \omega A_2
\end{equation}
with $\sigma_2^\dagger S \sigma_2=S^*$ and $\sigma_2^\dagger T \sigma_2=T^*$
and the two faithful complex representations
\begin{equation}
\MoreRep{2}{2}: S = A_1\qquad  T = \omega^2 A_2;\qquad \MoreRep{2}{3}: S = A_1,\qquad T = A_2
\end{equation}
with $\sigma_2^\dagger S_{2'} \sigma_2=S_{2''}^*$ and $\sigma_2^\dagger T_{2'} \sigma_2=T_{2''}^*$
where 
\begin{equation}
A_1=\frac{-1}{\sqrt{3}}\left(
\begin{array}{cc}
 i & {\tilde{\omega}} \sqrt{2} \\
- {\tilde{\omega}}^{-1} \sqrt{2} & -i
\end{array}
\right),
\qquad
A_2=\left(
\begin{array}{cc}
 \omega  & 0 \\
 0 & 1
\end{array}
\right)
\end{equation}
with ${\tilde{\omega}}=e^{2\pi\I /24}$. For all two-dimensional representations, we find the matrix
\begin{equation}
U=U_2\equiv\diag({\tilde{\omega}}^{-5}, {\tilde{\omega}}^5)
\end{equation}
which represents the automorphism $u$ via $U \rho(g)^*U^{-1}= \rho(u(g))$.
For the three-dimensional representation
\begin{equation}
\rho(S)=\frac{1}{3}\left(\begin{array}{ccc}
-1&2 \omega&2 \omega^2\\
2  \omega^2&-1&2  \omega\\
2  \omega&2  \omega^2&-1
\end{array}\right),\qquad \rho(T)=\left(\begin{array}{ccc}
1&0&0\\
0&\omega&0\\
0&0&\omega^2
\end{array}\right)
\end{equation}
the matrix U of Eq. (\ref{eq:def-gen-CP-U}) is given by $U=\rho(T)$ with again
$
U \rho(T)^* U^{-1}=\rho(T^2),\; U \rho(S)^* U=\rho(S^3)$, for the one dimensional representations we take $U=\rho(T)$ as for the three-dimensional representations.

In summary, we have thus found the one unique non-trivial outer automorphism (up to inner automorphisms) of $T^\prime$ and thus the unique CP transformation \footnote{This notation may lead to misinterpretations. What is meant is that for a collection of fields $\{ \varphi_i\}$ transforming as $\varphi_i\sim \Rep{r}_i$ under the group $G$, under CP each field has to transform in the way indicated below, up to an undetermined global phase. The transformation property of products of fields such as $\varphi^n\sim \Rep{r'}$ may vary up to a global phase, which is determined by the global phase of the CP transformation of $\varphi$. The CP properties of products are not determined by their transformation properties, but rather depend on the phase conventions adopted when defining the Clebsch-Gordon coefficients. It is possible to construct the Clebsch-Gordan coefficients such that the CP transformation properties are manifest for all covariants, sth.~we did not do here. To avoid confusion, we display invariants explicitly throughout and discuss CP on the level of the fields.}
\begin{equation}
\label{eq:TprimeCP}
\MoreRep{1}{i}\rightarrow \omega^{i-1} \MoreRep{1}{i}^*\qquad \MoreRep{2}{i}\rightarrow \diag({\tilde{\omega}}^{-5}, {\tilde{\omega}}^5)\MoreRep{2}{i}^*\qquad \Rep{3}\rightarrow \diag(1,\omega, \omega^2)\Rep{3}^*.
\end{equation}

Let us now use this insight to investigate a claim that there is geometrical CP violation in grand unified models based on $T^\prime$\cite{Chen:2009lr,Meroni:2012ty}. We consider the model discussed in~\cite{Chen:2009lr} and introduce $(T_1,T_2) \sim \MoreRep{2}{1}$ which transforms as $\Rep{10}$ of $\SU{5}$ and includes the first two generations of up-type quarks and the flavons $\phi\sim \Rep{3}$ and  $\phi^\prime \sim \Rep{3}$. Auxiliary $Z_{12} \times Z_{12}$ symmetries are introduced such that the one-two sector of the mass matrix is described by\footnote{We use the Clebsch-Gordan coefficients given in App.~A of \cite{Feruglio:2007yq} for the Kronecker products.} 
\begin{align}
-\mathcal{L}_{TT}&=y_c T T \phi^2 +y_u T T {\phi^\prime}^3+\hc\label{eq:invariantsChen}\\
&\equiv y_c  \frac32\frac{2-i}{2} (T T)_{\Rep{3}} (\phi^2)_{\Rep{3}} +y_u \frac13 [(T T)_{\Rep{3}} \phi^\prime]_{\MoreRep{1}{3}} ({\phi^\prime}^2)_{\MoreRep{1}{2}}+\hc\nonumber\\
&=y_c \frac32 \frac{2-i}{2} \left\{ (1-i)\,T_1 T_2\, (\phi_1^2-\phi_2\phi_3)+ i \,T_1^2 \left( \phi_2^2-\phi_1\phi_3\right)+ T_2^2 \left(\phi_3^2-\phi_1 \phi_2\right)
\right\}+\nonumber\\\nonumber
&+y_u \frac13 \left\{\left(2 \phi^\prime_1\phi^\prime_3+{\phi^\prime _2}^2\right) \left(i T_1^2 \phi^\prime_1+(1-i) T_1 T_2
   \phi^\prime_2+T_2^2 \phi^\prime_3\right)\right\}+\hc\;,
\end{align}
where we have omitted (Higgs-) fields that do not transform under the flavour symmetry and a suppression by some high-energy scale of a sufficient power to make $y_i$ dimensionless  is understood. 

It is assumed that the VEVs 
\begin{align}
\vev{\phi^\prime}=(1,1,1)V^\prime, \qquad \vev{\phi}=(0,0,1)V \qquad V,V^\prime\in \mathbbm{R}
\label{eq:chen-vev}
\end{align}
are real, which may be justified by a CP transformation. There is only one CP transformation\footnote{Note that this also determines the global phase of $U$.} left invariant, namely the one corresponding to the outer automorphism $u^\prime=\conj{T^2}\circ u$ represented on the three dimensional representation by the identity matrix 
\begin{equation}
\label{eq:defUprime}
\MoreRep{1}{i}\rightarrow \MoreRep{1}{i}^*\qquad \MoreRep{2}{i}\rightarrow \diag(\omega{\tilde{\omega}}^{-5}, \omega^{-1}{\tilde{\omega}}^5)\MoreRep{2}{i}^*\qquad \Rep{3}\rightarrow \Rep{3}^*.
\end{equation}
and therefore  $\braket{\phi^\prime}\rightarrow {\braket{\phi^\prime}}^*$ and $\braket{\phi}\rightarrow \braket{\phi}^*$.

This results in the following 1-2 block of the up-type quark mass matrix $M_u$:
\begin{equation}
y_u \left( \begin{array}{cc}
\I & \frac{{1-\I}}{2}\\
\frac{{1-\I}}{2}& 1
\end{array}\right){V^\prime}^3+y_c \left( \begin{array}{cc}
0& 0\\
0& 1-\frac{\I}{2}
\end{array}\right){V}^2\;.
\end{equation}
 At this point the parameters $y_{u,c}$ and VEVs are chosen real and it is claimed that the phases emerging from the complex Clebsch-Gordon coefficients explain CP violation. Therefore it is natural to ask whether this choice of parameters can be justified by a symmetry. The only candidate symmetry is a generalised CP symmetry of type \eqref{eq:CP-def}, which we explicitly state in \Eqref{eq:defUprime}. As we have shown how the various fields have to transform under the generalised CP symmetry we can now easily determine how the invariants of Eq. (\ref{eq:invariantsChen}) transform~\footnote{Note that inner automorphisms correspond to group transformations and therefore only outer automorphism can give non-trivial constraints when acting on group invariants. Here there is only one non-trivial outer automorphism(up to inner automorphisms). }:
 \begin{align}
 CP[T T \phi^2]=-\frac{4+3 \I}{5}  (T T \phi^2)^*\qquad CP[T T {\phi^\prime}^3]=-\I  (T T {\phi^\prime}^3)^*.
 \end{align} 
Therefore invariance under CP requires $\arg(y_c)=-\frac{1}{2}\arg (-4-3 \I)=-\frac{1}{2}\arctan \frac{3}{4}$ and $\arg y_u=\frac{\pi}{4}$ and the generalised CP \eqref{eq:defUprime} is explicitly broken by real couplings $y_u, y_c$, which was assumed in Ref.~\cite{Chen:2009lr}. Note that also the relative phase between the two couplings does not agree with 'geometrical' CP violation. This also shows that the results obtained in Ref. \cite{Chen:2009lr} are completely basis dependent and therefore unphysical.

Although the VEVs \eqref{eq:chen-vev}  are invariant under the generalised CP transformation \eqref{eq:defUprime}, in the full model~\cite{Chen:2009lr} there are additional scalar fields e.g. $\psi\sim \MoreRep{2}{2}$ with $\vev{\psi}\sim(1,0)$ which are not invariant under the generalised CP transformation \eqref{eq:defUprime}. Hence, if the phases of the couplings are changed in accordance with the consistent CP transformation \eqref{eq:defUprime}, CP will be broken spontaneously. Obviously, all predictions depend on the VEV alignment. In Ref.~\cite{Chen:2009lr}, no dynamical mechanism was given to generate the VEV configuration.

Different invariants were used in the other grand unified $T^\prime$ model~\cite{Meroni:2012ty} claiming a geometric origin of CP violation. In the following, we argue that the CP phases in this model do not have a geometric origin as well. The argument is done in two steps: 1) We choose a CP transformation, which is not broken by the VEVs. 2) CP is explicitly broken by two different couplings in the superpotential. 

1) As the CP transformation defined in \Eqref{eq:defUprime} is not broken by real VEVs of the singlet and triplet flavons, it is enough to consider the four doublets $\psi^{\prime(\prime)}=(\psi^{\prime(\prime)}_1,\psi^{\prime(\prime)}_2)^T$ and $\tilde\psi^{\prime(\prime)}=(\tilde\psi^{\prime(\prime)}_1,\tilde\psi^{\prime(\prime)}_2)^T$, which obtain the VEVs
\begin{align}\label{eq:TprimeVEValign}
\braket{\psi^\prime} &= \begin{pmatrix}1\\0\end{pmatrix} \psi_0^\prime,&
\braket{\tilde\psi^\prime}&= \begin{pmatrix}1\\0\end{pmatrix}\tilde\psi_0^{\prime}, &
\braket{\psi^{\prime\prime}}  &= \begin{pmatrix}0\\1\end{pmatrix} \psi_0^{\prime\prime}, &
\braket{\tilde\psi^{\prime\prime}}&= \begin{pmatrix}0\\1\end{pmatrix}\tilde \psi_0^{\prime\prime}\;.
\end{align}
We modify the CP transformation \eqref{eq:defUprime} by a phase redefinition of the doublet fields
\begin{align}\label{eq:defUprimePhase}
\psi^\prime&\rightarrow \diag(1, \omega^{-2}{\tilde{\omega}}^{10})\psi^{\prime*} &
\psi^{\prime\prime}&\rightarrow \diag(\omega^{2}{\tilde{\omega}}^{-10},1)\psi^{\prime\prime*} \\\nonumber
\tilde\psi^\prime&\rightarrow \diag(1, \omega^{-2}{\tilde{\omega}}^{10})\tilde\psi^{\prime*} &
\tilde\psi^{\prime\prime}&\rightarrow \diag(\omega^{2}{\tilde{\omega}}^{-10},1)\tilde\psi^{\prime\prime*} 
\end{align}
such that the VEVs do not break this CP transformation.

2) Using this CP transformation, we consider two couplings in the superpotential\footnote{A CP transformation relates the (holomorphic) superpotential with the anti-holomorphic superpotential. Similarly to \Eqref{eq:invariantsChen}, we omit Higgs fields that do not transform under the flavour symmetry and a suppression by some high-energy scale $\Lambda$ of a sufficient power to make $y_i$ dimensionless is understood. We use the Clebsch-Gordan coefficients given in Tab.~7 of Ref.~\cite{Meroni:2012ty} itself. Note that the second operator is only defined up to a sign in Ref.~\cite{Meroni:2012ty}. However, this does not affect the discussion.} 
\begin{align}
\mathcal{W}_{Y_u}&
\supset
 y_{22} TT\tilde\psi^{\prime\prime2}\tilde\zeta^\prime
+ y_{21} TT\tilde\phi\tilde\psi^{\prime2}\tilde\zeta^\prime\\\nonumber
&\equiv
 y_{22} ((T \tilde \psi^{\prime\prime})_{\Rep{3}} (T \tilde \psi^{\prime\prime})_{\Rep{3}})_{\MoreRep{1}{3}} \tilde \zeta^\prime +
 y_{21} ((T \tilde \phi)_{\MoreRep{2}{2}} \tilde \zeta^\prime)_{\MoreRep{2}{3}} (\tilde\psi^\prime(T \tilde \psi^{\prime})_{\Rep{3}})_{\MoreRep{2}{2}}\\
&=\frac{y_{22}}{\sqrt{3}}  \tilde\zeta^\prime\left\{
T_2^2 \tilde\psi_2^{\prime\prime2} +(1+\I) T_1 \tilde\psi^{\prime\prime}_1(T_1\tilde\psi_2^{\prime\prime}+T_2\tilde\psi_1^{\prime\prime})
\right\}
\nonumber\\\nonumber
&+\frac{y_{21}}{3\sqrt{2}} \tilde\zeta^\prime \left\{
T_1^2 \left[(1-\I)\tilde\phi_1 \tilde\psi_2^{\prime2}-(1+\I)\tilde\phi_2\tilde\psi_2^{\prime2}\right]
+T_2^2 \left[(1+\I)\tilde\phi_3 \tilde\psi_2^{\prime2}-2\tilde\phi_2\tilde\psi_1^{\prime}\tilde\psi_2^\prime\right]
\right.\\\nonumber
&\qquad\qquad\left.
+2\,T_1 T_2\left[(1-\I)\tilde\phi_1\tilde\psi_1^\prime\tilde\psi_2^\prime-\I\tilde\phi_3\tilde\psi_1^{\prime2}
\right]
\right\}
\end{align}
contributing to the 1-2 sector of the up-type quark mass matrix, where $T=(T_1,T_2)^T\sim\MoreRep{2}{1}$, $\tilde\zeta^\prime\sim\MoreRep{1}{2}$ and $\tilde\phi=(\tilde\phi_1,\tilde\phi_2,\tilde\phi_3)^T\sim\Rep{3}$. Similarly to the argument for \cite{Chen:2009lr}, we can now easily determine how the invariants transform under the generalised CP transformation
\begin{align}
CP[TT\tilde\psi^{\prime\prime2}\tilde\zeta^\prime] & =-\I (TT\tilde\psi^{\prime\prime2}\tilde\zeta^\prime)^* &
CP[TT\tilde\phi\tilde\psi^{\prime2}\tilde\zeta^\prime] & =- (TT\tilde\phi\tilde\psi^{\prime2}\tilde\zeta^\prime)^*\;.
\end{align}
Hence, CP is explicitly broken by choosing $y_{22}$ and $y_{21}$ real. As there is a relative phase difference between the two operators, it is not possible to redefine the CP transformation of $T$, such that there is no explicit CP violation.

The authors additionally propose a way to obtain the VEV alignment using driving fields using the method introduced in Ref.~\cite{Antusch:2011sx}. Let us analyse the flavon potential in more detail using the generalised CP transformation \eqref{eq:defUprime}, i.e. without the modification in \Eqref{eq:defUprimePhase}. It seems plausible to get real VEVs for the triplets and singlet fields, as these are eigenstates of CP. We will therefore concentrate on the doublets $\psi^{\prime(\prime)}$ and $\tilde\psi^{\prime(\prime)}$. 
The generalised CP transformation \eqref{eq:defUprime} fixes the phase (modulo $\pi$) of all couplings  and in particular\footnote{We dropped the fields $\epsilon_i$, which are singlets of $T^\prime$ and are required to adjust the charges of the shaping symmetries.}
\begin{align}
\mathcal{W}_f&\supset D_\psi (\lambda (\psi^{\prime\prime})^2 + \kappa\phi\zeta^\prime) \\\nonumber
&= \frac{1}{\sqrt{3}} \left[D_{\psi1} (\lambda \psi_2^{\prime\prime2} +\kappa \phi_3\zeta^\prime)
+D_{\psi2} (\I\lambda \psi_1^{\prime\prime2} +\kappa\phi_2\zeta^\prime )
+D_{\psi3} ((1-\I)\lambda \psi_1^{\prime\prime} \psi_2^{\prime\prime} +\kappa \phi_1\zeta^\prime)
\right]
\end{align}
where $D_\psi\sim\Rep{3}$ is a driving field, $\phi\sim\Rep{3}$ and $\zeta^\prime\sim\MoreRep{1}{2}$. CP invariance with respect to \Eqref{eq:defUprime} requires $\kappa$ to be real and $\arg(\lambda)=\pi/4$. Assuming the VEV alignment \eqref{eq:TprimeVEValign}, the F-term equation
\begin{equation}
0\stackrel{!}{=}\frac{\partial\mathcal{W}_f}{\partial D_{\psi1}}= \frac{1}{\sqrt{3}} (\lambda \psi_2^{\prime\prime2} +\kappa \phi_3\zeta^\prime)
\end{equation}
 leads to a complex VEV for $\psi^{\prime\prime}$ with  $\arg(\psi_0^{\prime\prime})=7\pi/8+ \mathbb{Z}\, \pi$  for $\lambda\kappa<0$, which conserve CP, and $3\pi/8+ \mathbb{Z}\, \pi$ for $\lambda\kappa>0$, which preserves $\psi_0^{\prime\prime}\to-CP[\psi_0^{\prime\prime}]$, a different CP transformation, which can also be extended to a symmetry of the full theory by changing the CP transformation of the doublets to $CP^\prime: \MoreRep{2}{i}\to -CP[\MoreRep{2}{i}]$.
Hence, it is not possible to break both CP transformations by the VEV of $\psi^{\prime\prime}$ alone. However, the VEV of a second doublet can break the remaining CP transformation, but the phases of the VEVs depend on a discrete choice.
The other doublet VEVs are related to the VEV of $\psi^{\prime\prime}$ via $\arg(\tilde\psi_0^{\prime\prime})=\arg(\psi_0^{\prime\prime}) +\mathbb{Z}\, \pi/2$ and $\arg(\psi_0^\prime),\arg(\tilde\psi_0^\prime)=-\arg(\psi_0^{\prime\prime}) +\mathbb{Z}\, \pi/2$, where the shift $\mathbb{Z}\,\pi/2$ depends on the sign of the respective couplings. Hence, there is a discrete set of phases of the VEVs. In analogy to the VEV of $\psi^{\prime\prime}$, each VEV breaks one of the two CP transformations.  Concluding, as soon as there are two VEVs, which break different CP transformations, it is possible to have CP violation.

It might be instructive to look at the potential for one doublet field $\psi\sim \MoreRep{2}{2}$ and study the VEV configurations that can be obtained in order to see if it is possible to obtain a phase prediction from a spontaneous breaking of the generalised CP. On renormalizable level there is only one coupling that depends on phases 
\begin{equation}
\lambda\frac{{\tilde{\omega}}^2}{\sqrt{3}}\left( \psi_1(\psi_1^3-(2-2 \I) \psi_2^3)\right)+\hc,
\end{equation}
where the phases have been adjusted such, that CP forces $\lambda$ to be real. We will focus on VEVs of the form $\vev{\psi}= (V e^ {\I \alpha},0)^T$ with $V>0$ that conserve the $Z_3$ subgroup generated by $T$. For $\lambda<0$ we find the minima $\{1,\I, -1,-\I\}(e^{\I \pi 11/24 },0)^T$, which conserve $\psi \rightarrow \{1,-1,1,-1\}CP[\psi] $, and for  $\lambda>0$ we find the minima $\{1,\I, -1,-\I\}(e^{\I \pi 5/24} ,0)^T$  which conserve $\psi \rightarrow -\I \{1,-1,1,-1\}CP[\psi] $. The additional solutions are due to fact that the phase dependent part of the potential has an accidental $Z_4$ symmetry $\psi\rightarrow \I \psi$, which will most likely be broken in a full theory such that one would expect only the CP conserving solutions to survive. The required real VEVs cannot be obtained in this simple setup. Note that if the VEV of $\psi$ conserves CP, the phases of the VEVs and of the couplings conspire that there is no CP violation, as shown in Section \ref{sec:physicalPhases}.

Let us briefly summarise our view on geometrical CP violation in $T^\prime$. To be able to talk about CP violation one has to apply the consistent CP symmetry of \Eqref{eq:defUprime} on the Lagrangian level. This will then fix the phase (modulo $\pi$) of most couplings. The phases of invariants, which are CP self-conjugate are not fixed by CP. In supersymmetric theories, the phases of all couplings in the superpotential are fixed (modulo $\pi$), because CP relates the superpotential with the anti-holomorphic superpotential. One could thus imagine a setup along the lines of ~\cite{Chen:2009lr,Meroni:2012ty} where this has been implemented and therefore the only source of CP violation are the VEVs of the doublet scalar fields, which break CP spontaneously. However, the phases of the VEVs are only determined up to a finite discrete choice.

For usual spontaneous breaking of CP one would expect the phases of the fields to depend on potential parameters and therefore not be determined by the group symmetry structure. 
The only way to get 'calculable phases', i.e.~phases that do not depend on potential parameters, seems to be if this CP breaking vacua is connected to an additional (accidental) CP symmetry of the potential as is the case for $\Delta(27)$ (see \Secref{sec:Delta27}). For $T'$, however, there cannot be such an additional generalised CP besides the CP transformations which are connected to the unique non-trivial CP transformation by some group transformation, since the outer automorphism group is $Z_2$.

\mathversion{bold}
\subsection{$\Delta(27)\cong(Z_3\times Z_3)\rtimes Z_3\cong\SG{27}{3}$}
\mathversion{normal}
\label{sec:Delta27}
The group $\Delta(27)=\braket{A,B \vert A^3=B^3=(AB)^3=E}$~\footnote{$\Delta(27)$ has been first used in the lepton sector in  \cite{de-Medeiros-Varzielas:2007gf}.} is another interesting group from the standpoint of CP violation. Its automorphism structure is quite involved. The centre of the group is isomorphic to $Z_3$ and generated by the group element $X=A^2BAB^2$ with $\conj{X}=\id$ and the inner automorphism group has the structure $ Z_3\times Z_3$. The outer automorphism group is generated by 
\begin{align}
 u_1: (A,B) &\to(ABA^2,B^2AB)\;,&
 u_2:(A,B)&\to(ABAB,B^2)\;.
\end{align}
It is isomorphic to $\mathrm{GL}(2,3)$, i.e. the general linear group of $2\times2$ matrices over the field $Z_3$. The multitude of outer automorphisms can be traced back to the various symmetries of the character table shown in \Tabref{tab:ctbDelta27} that are due to the fact that there are so many one-dimensional representations. Together with the inner automorphisms these generators generate the full automorphism group, which is of order $432$. In summary the automorphism structure presents itself as:
\setcounter{nodecount}{0}
\begin{table}
\centering
\begin{tabular}{c|ccccccccccc}
&\tabnode{$E$}&\tabnode{$BABA$}& \tabnode{$ABA$} &\tabnode{$A$}&\tabnode{$BAB$}&\tabnode{$AB$}&\tabnode{$A^2$}&\tabnode{$B^2$}&\tabnode{$B$}&\tabnode{$BA^2BAB$}&\tabnode{$AB^2ABA$} \\ \hline
$\MoreRep{1}{1}$&     1 & 1&   1 & 1&  1&  1&  1 & 1&  1 & 1 & 1\\
\tabnode{$\MoreRep{1}{2}$}&     1&  $\omega$ & $\omega^2$ & 1 & $\omega$ & $\omega^2$& 1 & $\omega$ & $\omega^2$ & 1&  1\\
\tabnode{$\MoreRep{1}{3}$}&      1 &$\omega^2$ & $\omega$ &  1 &$\omega^2$& $\omega$ & 1 &$\omega^2$& $\omega$ & 1 & 1\\
\tabnode{$\MoreRep{1}{4}$}&     1  &$\omega$ &  $\omega$ & $\omega^2$ &$\omega^2$ &$\omega^2$ & $\omega$ & 1&  1&  1&  1\\
\tabnode{$\MoreRep{1}{5}$}&     1 &$\omega^2$ & 1& $\omega^2$ & 1&  $\omega$ & $\omega$ & $\omega$ &$\omega^2$& 1 & 1\\
\tabnode{$\MoreRep{1}{6}$}&     1  &1 &$\omega^2$ &$\omega^2$ & $\omega$ & 1&  $\omega$ & $\omega^2$ & $\omega$ & 1&  1\\
\tabnode{$\MoreRep{1}{7}$}&      1& $\omega^2$ &$\omega^2$ & $\omega$ & $\omega$ &  $\omega$ &$\omega^2$ & 1&  1 & 1&  1\\
\tabnode{$\MoreRep{1}{8}$}&     1 & 1 & $\omega$ & $\omega$ &$\omega^2$& 1 &$\omega^2$& $\omega$ & $\omega^2$ & 1 & 1\\
\tabnode{$\MoreRep{1}{9}$}&      1 & $\omega$ & 1 & $\omega$ & 1 &$\omega^2$ &$\omega^2$ &$\omega^2$& $\omega$ & 1 & 1\\
\tabnode{$\Rep{3}$}&     3&  .&  .&  .&  .&  .&  .&  .&  .&  $3 \omega$ &$3\omega^2$\\
\tabnode{$\Rep{3}^*$}&    3  &.&  .&  .&  .&  . & .&  . & .& $3 \omega^2$&  $3 \omega$\\
\end{tabular}
\caption{Character table of $\Delta(27)$. The first line indicates
  representatives of the different conjugacy classes. Zeroes in the
  character table are denoted by a dot $.$  and $\omega$ is the third
  root of unity $\omega=e^{2\pi\I/3}$. The arrows illustrate the generators of the outer automorphism group $u_1$(blue) and $u_2$(red).\label{tab:ctbDelta27}}
\begin{tikzpicture}[overlay]
\node [above=.3cm,minimum width=0pt] at (1) (c1){};
\node [above=.3cm,minimum width=0pt] at (2) (c2){};
\node [above=.3cm,minimum width=0pt] at (3) (c3){};
\node [above=.3cm,minimum width=0pt] at (4) (c4){};
\node [above=.3cm,minimum width=0pt] at (5) (c5){};
\node [above=.3cm,minimum width=0pt] at (6) (c6){};
\node [above=.3cm,minimum width=0pt] at (7) (c7){};
\node [above=.3cm,minimum width=0pt] at (8) (c8){};
\node [above=.3cm,minimum width=0pt] at (9) (c9){};
\node [above=.3cm,minimum width=0pt] at (10) (c10){};
\node [above=.3cm,minimum width=0pt] at (11) (c11){};
\node [left=.2cm,minimum width=0pt] at (12) (r12){};
\node [left=.2cm,minimum width=0pt] at (13) (r13){};
\node [left=.2cm,minimum width=0pt] at (14) (r14){};
\node [left=.2cm,minimum width=0pt] at (15) (r15){};
\node [left=.2cm,minimum width=0pt] at (16) (r16){};
\node [left=.2cm,minimum width=0pt] at (17) (r17){};
\node [left=.2cm,minimum width=0pt] at (18) (r18){};
\node [left=.2cm,minimum width=0pt] at (19) (r19){};
\node [left=.2cm,minimum width=0pt] at (20) (r31){};
\node [left=.2cm,minimum width=0pt] at (21) (r32){};
\node [right=.2cm,minimum width=0pt] at (12) (r12ri){};
\node [right=.2cm,minimum width=0pt] at (13) (r13ri){};
\node [right=.2cm,minimum width=0pt] at (14) (r14ri){};
\node [right=.2cm,minimum width=0pt] at (15) (r15ri){};
\node [right=.2cm,minimum width=0pt] at (16) (r16ri){};
\node [right=.2cm,minimum width=0pt] at (17) (r17ri){};
\node [right=.2cm,minimum width=0pt] at (18) (r18ri){};
\node [right=.2cm,minimum width=0pt] at (19) (r19ri){};
\node [right=.2cm,minimum width=0pt] at (20) (r31ri){};
\node [right=.2cm,minimum width=0pt] at (21) (r32ri){};
\draw [<->,out=45,in=135,blue!50, very thick,below=1cm] (c3) to (c5);
\draw [<->,out=45,in=135,blue!50, very thick,below=1cm] (c4) to (c9);
\draw [<->,out=45,in=135,blue!50, very thick,below=1cm] (c7) to (c8);
\draw [<->,out=45,in=135,blue!50, very thick,below=1cm] (c10) to (c11);
\draw [->,out=45,in=135,red!50, very thick,below=1cm] (c2) to (c5);
\draw [->,out=10,in=170,red!50, very thick,below=1cm] (c3) to (c4);
\draw [<-,out=90,in=90,red!50, very thick,below=1cm] (c2) to (c4);
\draw [->,out=45,in=90,red!50, very thick,below=1cm] (c5) to (c7);
\draw [<-,out=90,in=135,red!50, very thick,below=1cm] (c3) to (c6);
\draw [<-,out=45,in=135,red!50, very thick,below=1cm] (c6) to (c7);
\draw [<->,out=45,in=135,red!50, very thick,below=1cm] (c8) to (c9);
\draw [<->,out=-45,in=45,blue!50, very thick,below=1cm] (r31ri) to (r32ri);
\draw [<->,out=-45,in=45,blue!50, very thick,below=1cm] (r12ri) to (r14ri);
\draw [<->,out=-45,in=45,blue!50, very thick,below=1cm] (r13ri) to (r17ri);
\draw [<->,out=-45,in=45,blue!50, very thick,below=1cm] (r16ri) to (r18ri);
\draw [->,out=225,in=135,red!50, very thick,below=1cm] (r12) to (r19);
\draw [<-,out=-90,in=-90,red!50, very thick,below=1cm] (r18) to (r19.north);
\draw [<-,out=225,in=135,red!50, very thick,below=1cm] (r13) to (r18);
\draw [->,out=90,in=100,red!50, very thick] (r13.south) to (r15);
\draw [->,out=225,in=135,red!50, very thick,below=1cm] (r15) to (r16);
\draw [<-,out=180,in=180,red!50, very thick,below=1cm] (r12) to (r16);
\end{tikzpicture}
\end{table}
\begin{align}
\zentrum{\Delta(27)}&\cong Z_3&
\aut{\Delta(27)}&\cong (((Z_3 \times Z_3) \rtimes Q_8) \rtimes Z_3) \rtimes Z_2\\\nonumber
\inn{\Delta(27)}&\cong Z_3\times Z_3&
\out{\Delta(27)}&\cong \mathrm{GL}(2,3)\;.
\end{align}
The outer automorphism $u_1$ acts on the representations as
\begin{equation}
\MoreRep{1}{2}\leftrightarrow\MoreRep{1}{4},\qquad\MoreRep{1}{3}\leftrightarrow\MoreRep{1}{7},\qquad\MoreRep{1}{6}\leftrightarrow\MoreRep{1}{8},\qquad \Rep{3}\leftrightarrow\Rep{3}^*
\end{equation}
where e.g.  $\MoreRep{1}{2}\rightarrow\MoreRep{1}{4}$ is to be read as $ \rho_{\MoreRep{1}{4}}= \rho_{\MoreRep{1}{2}}\circ u_1$ etc., and the outer automorphism $u_2$ acts as
\begin{equation}
\MoreRep{1}{2}\rightarrow \MoreRep{1}{9}\rightarrow \MoreRep{1}{8}\rightarrow \MoreRep{1}{3}\rightarrow \MoreRep{1}{5}\rightarrow \MoreRep{1}{6}\rightarrow \MoreRep{1}{2}
\end{equation} 
From this it is trivial to determine the representations of the automorphisms for the one-dimensional representations. Let us therefore focus on the three dimensional representation  $\Rep{3}$ generated by 
\begin{equation}
\rho(A)=T_3, \qquad \rho(B)=\diag(1,\omega, \omega^2).
\end{equation}
The two generators of the outer automorphism group act on $\phi\sim(\Rep{3},\Rep{3}^*)$ as
\begin{align}
U(u_1)=\left( \begin{array}{cc} \tilde U&0\\ 0&\tilde U^*  \end{array}\right) \quad \mathrm{with}\quad \tilde U=\frac{1}{\sqrt{3}}\left( \begin{array}{ccc} \omega ^2&\omega&1\\ \omega &\omega ^2&1\\ 1& 1&1   \end{array}\right)
\end{align}
and 
\begin{align}
U(u_2)=\left( \begin{array}{cc} 0&\tilde U\\ \tilde U^*&0  \end{array}\right) \quad \mathrm{with}\quad \tilde U=\left( \begin{array}{ccc} \omega^2 & 0&0 \\0&0&\omega \\ 0&\omega^2&0 \end{array}\right).
\end{align}
All automorphisms can be generated from the generators $u_i$ by composition and the representation matrices $U(\mathrm{aut})$ may be obtained with the help of \Eqref{eq:group-structure-aut}. We have therefore found a complete classification of possible CP transformations that may be implemented in a model based on $\Delta(27).$ There are 48 outer automorphisms generated by $u_1$ and $u_2$ that may in principle give physically distinct CP transformations with distinct physical implications, however as a model that is invariant under CP will also be invariant under $\mathrm{CP}^n$ it is sufficient to consider which subgroups of the automorphism groups is realised. 

It is instructive to look at some of these subgroups in detail. Let us for example consider the CP transformation $\phi \rightarrow \phi^*$ or $U(h_1)=\mathbbm{1}_3$ that corresponds to the outer automorphism $h_1: (A,B)\to(A,B^2)$, which can be expressed in terms of the generators as $h_1=u_1\circ u_2^2\circ u_1^{-1}\circ u_2\circ u_1^{-1}\circ u_2^{-1}\circ u_1^{-1}\circ \conj{A}^{-1}\circ u_1^{-1}$. This outer automorphism squares to one and therefore generates a $Z_2$ subgroup of the automorphism group. Contrary to the situation we have encountered before, where the outer automorphism group was a $Z_2$, this is not the only solution.
As a further example we may consider the $Z_2$ subgroup generated by $u_1\circ u_2^2\circ u_1^{-1}\circ u_2\circ u_1^{-1}\circ u_2^{-2}$ with $h_2:(A,B)\to(ABA,B)$ which according to \Eqref{eq:group-structure-aut} is represented by
\begin{align}
U(h_2)=\left( \begin{array}{ccc} \omega & 0&0 \\0&0&1 \\ 0&1&0 \end{array}\right).  
\label{eq:Z2subautdelta27}
\end{align}
We will use this matrix later on. 
Let us now use this machinery to tackle a physical question, namely the so-called geometrical CP violation. 'Geometrical' CP-violation~\cite{Branco:1983tn,*deMedeirosVarzielas:2011zw,*de-Medeiros-Varzielas:2012fk,*Varzielas:2012pd,*Bhattacharyya:2012pi} denotes the following: If one considers a triplet of Higgs doublets $H=(H_1, H_2, H_3)\sim \Rep{3}$ the only phase dependent term in the scalar potential is given by
\begin{equation}
I \equiv  \sum_{i\neq j \neq k} (H_i^\dagger H_j)  (H_i^\dagger H_k). 
\end{equation}
Let us now investigate how the term transforms under the two generators $u_1$ and $u_2$ of the outer automorphism group. We find
\begin{equation}
CP_{u_1}[I]=-\frac{1}{3} I^* +\frac{2}{3}I +\sum_i \frac{1}{3}(H_i^\dagger H_i)^2+\sum_{i\neq j} (H_i^\dagger H_i)(H_j^\dagger H_j),\qquad CP_{u_2}[I]=\omega^2 I
\end{equation}
and we thus find the invariant combinations 
\begin{equation}
CP_{u_1}[I-I^*]=I-I^*\qquad CP_{u_2^3}[I]=I
\end{equation}
Clearly invariance under $u_1$ requires further non-trivial relations among the other couplings in the scalar potential which do not depend on phases and thus do not concern us here. 

Let us investigate the case where the theory is invariant under $h_1$ which corresponds to the 'usual' CP transformation $\phi\rightarrow \phi^*$ and forces the coupling $\lambda_4$ multiplying $I$ to be real. For $\lambda_4<0$ one finds the global minimum 
\begin{equation}
\ev{H}=\frac{v}{\sqrt{3}} (1,\omega,\omega^2)
\end{equation}
and for $\lambda_4>0$ one finds
\begin{equation}
\ev{H}=\frac{v}{\sqrt{3}} (\omega^2,1,1).
\end{equation}
Both VEV configurations correspond to generalised CP transformations $H\rightarrow U H^*.$ For $\lambda_4<0$ it is for example given by $U=\rho(B^2)$ which is clearly part of $\Delta(27)$ and therefore up to an inner automorphism corresponds to $h_1$. The phases of the VEVs thus do not imply spontaneous CP violation. For $\lambda_4>0$ the VEV configuration leaves the CP transformation corresponding to the outer automorphism $h_2$ given in Eq. (\ref{eq:Z2subautdelta27}) invariant. However, there is something that is much harder to understand about this VEV configuration: the generalised CP symmetry corresponding to this configuration is not a symmetry of the Lagrangian. It would be a symmetry if the phase of $\lambda_4$ would be the same as $\omega$, as $CP_{h_2}[I]=\omega I^*$.
So here we are confronted with the puzzling situation where a VEV configuration is more symmetric than the original Lagrangian. This is also denoted as calculable phases. 
 
This conundrum can be solved if there is a generalised CP trafo that is left invariant by the VEV and is compatible with $\lambda_4$ being real. Since we have a complete classification of all generalised CP transformations we can answer this question and indeed we find the CP transformation 
\begin{align}
\left(\begin{array}{c}H\\H^*\end{array}\right)=U \left(\begin{array}{c}H^*\\H\end{array}\right) \quad\mathrm{with}\quad
 U=\left(\begin{array}{cc}0&\tilde{U}\\\tilde{U}^*&0\end{array}\right), \qquad \tilde{U}=\left(
\begin{array}{ccc}
 0 & 0 & \omega^2 \\
 0 & 1 & 0 \\
\omega & 0 & 0
\end{array}
\right)
\end{align}
which represents the outer automorphism $u:(A,B)\to(A B^2 A B, A B^2 A^2)$ via Eq. \eqref{eq:aut-def-from-U}, where $u=u_2^3\circ \mathrm{conj}(A)$ and that gives 
\begin{equation}
CP_u[\vev{H}]=\vev{H} \qquad \mathrm{for }\qquad \ev{H}=\frac{v}{\sqrt{3}} (\omega^2,1,1), \qquad CP_u[I]=I
\end{equation}
Note that this CP transformation acts as $H\rightarrow \tilde{U} H$, which is not something you would naively expect, but it is an outer automorphism and therefore it is justified to call it a CP transformation. Furthermore, this becomes apparent when one looks at how the outer automorphism $u$ acts on representations. It interchanges the one-dimensional representations
\begin{equation}
\MoreRep{1}{2}\leftrightarrow \MoreRep{1}{3},\qquad \MoreRep{1}{5}\leftrightarrow \MoreRep{1}{9},\qquad \MoreRep{1}{6}\leftrightarrow \MoreRep{1}{8},
\end{equation} 
making the "CP-character" of the transformation more apparent. 
An alternative independent explanation of geometric CP violation has been given in Ref.~\cite{Degee:2012sk}.
\mathversion{bold}
\subsection{$Z_9\rtimes Z_3\cong\SG{27}{4}$}
\mathversion{normal}
Similarly to $\Delta(27)$, the group $Z_9\rtimes Z_3=\SG{27}{4}=\braket{A,B|A^9=B^3=BAB^{2}A^2=E}$\footnote{The possibility of having $Z_9\rtimes Z_3$ as a flavour group in the lepton sector has been first mentioned in Ref.~\cite{Parattu:2010cy}.} has a more complicated automorphism group structure. The group is the semi-direct product of $Z_9$ generated by $A$ (with $A^9=E$) with $Z_3$ generated by $B$ (with $B^3=E$) defined by $BAB^{-1}=A^7$. The centre of the group is isomorphic to $Z_3$ and generated by $A^3$. Hence, the inner automorphism group has the structure $Z_3\times Z_3$. The outer automorphism group is generated by 
\begin{align}
u_{1}:&(A,B)\to(AB,B^2 A^6 B^2 A^3)\\\nonumber
u_2:&(A,B)\to(AB^4AB^4A^6,B^2 A^6 B^2 A^6)\;.
\end{align}
and the structure of the automorphism group may be summarised as
\begin{align}
\zentrum{G}&\cong Z_3&
\aut{G}&\cong ((Z_3 \times Z_3) \rtimes Z_3) \rtimes Z_2\\\nonumber
\inn{G}&\cong Z_3\times Z_3&
\out{G}&\cong S_3\;.
\end{align}
There is a faithful three dimensional representation given by
\begin{equation}
\rho(A)=\left(\begin{array}{ccc}
0& 1& 0\\ 0& 0& \omega^2 \\ \omega^2& 0& 0 
\end{array}\right),
\qquad
\rho(B)=\left(\begin{array}{ccc}
\omega^2& 0& 0\\ 0& 1& 0 \\ 0& 0& \omega 
\end{array}\right).
\end{equation}
The generators of the outer automorphisms can be obtained in the same way as before and act on $(\Rep{3},\Rep{3}^*)$ as
\begin{align}
U(u_1)=\left(\begin{array}{cc}0&\tilde{U} \\ \tilde{U}^*&0 \end{array} \right) \qquad \mathrm{with } \qquad \tilde{U}=\diag(1,1,\omega^2)
\end{align}
and 
\begin{align}
U(u_2)=\left(\begin{array}{cc}\tilde{U}&0 \\ 0&\tilde{U}^* \end{array} \right) \qquad \mathrm{with } \qquad \tilde{U}=\left(
\begin{array}{ccc}
 0 & 1 & 0 \\
 1 & 0 & 0 \\
0 & 0 & \omega^2
\end{array}
\right).
\end{align}
\mathversion{bold}
\subsection{$Q_8\rtimes A_4\cong\SG{96}{204}$}
\mathversion{normal}
\label{sec:CP-Q8}
Let us also consider our favourite group, $Q_8\rtimes A_4$~\cite{Holthausen:2011vd,Holthausen:2012wz}  generated by $S,T,X$ with
\begin{equation}
 S^2=T^3=X^4=(ST)^3=SXSX^{3}=T^{2}X (T^{2}X^{3})^2
= (STX^{3}T^{2})^2  =E
\end{equation}
the smallest group that may realise the VEV alignment. Its centre is given by $\zentrum{Q_8\rtimes A_4}=\{ E,X^2\}\cong Z_2$ and its outer automorphism group is generated by 
\begin{align}
h_4:&(S,T,X)\to(S,T^2,SX),&
h_5:&(S,T,X)\to(S,T^2,X^3), \nonumber\\ 
h_6:&(S,T,X)\to(S T^2 S T X^3,T,T^2 X T).
\label{eq:Q8aut-gen}
\end{align}
These generators act on the character table and representations in the way indicated in \Tabref{tab:ctblQ8A4}. Together with the inner automorphisms, the automorphism group is of order 576 and its structure may be summarised as:
\begin{align}
\zentrum{Q_8\rtimes A_4}&\cong Z_2&
\aut{Q_8\rtimes A_4}&\cong((A_4 \times A_4) \rtimes Z_2) \rtimes Z_2\\\nonumber
\inn{Q_8\rtimes A_4}&\cong Z_2^4 \rtimes Z_3&
\out{Q_8\rtimes A_4}&\cong D_{12}\;.
\end{align}
Let us discuss how the generators of the automorphism group may be represented upon the vector
\begin{equation} 
\phi=\left( \begin{array}{c} \varphi_C\\ \varphi_C^* \end{array}\right)  
\end{equation}
with $\varphi_C\sim \MoreRep{4}{2}$ upon which the group generators act as
\begin{equation}
\rho(S)=\left(
\begin{array}{cc}
S_4&0\\0&S_4
\end{array}
\right),\qquad
\rho(T)=\left(
\begin{array}{cc}
\omega^2 T_4&0\\0&\omega T_4
\end{array}
\right),\qquad \mathrm{and} \qquad 
\rho(X)=\left(
\begin{array}{cc}
X_4&0\\0&X_4
\end{array}
\right)
\end{equation}
with 
\begin{align}
S_4\equiv&
\sigma_3 \otimes\sigma_1,
&
T_4&\equiv \diag(T_3,1),
&X_4\equiv&
-\I \sigma_2 \otimes\sigma_3
\end{align}
and $\rho(S,X)^*=\rho(S,X )$ but $\rho(T)^*\notin \im \rho$. One solution to Eq. (\ref{eq:CP-def}) is the analogue of the $A_4$ case, $U=\diag(U_4,U_4)$ with $U_4=\diag(\tilde{U}_3\equiv T_3 U_3 T_3^{-1}, 1)$. The matrix $U_3$ has been defined in \Eqref{eq:defU3}. This generator  acts on the generators of the group as
\begin{align}
U \rho(S)^* U^{-1}=\rho(S),\qquad U \rho(T)^* U^{-1}=\rho(T^2),\qquad U \rho(X)^* U^{-1}=\rho(SX)
\label{eq:aut-def}
\end{align}
and therefore represents the automorphism $h_4$. Before discussing other solutions to Eq. (\ref{eq:CP-def}), let us demonstrate how this outer automorphism can be represented for the other representations. For the representation $\MoreRep{4}{1}$ we find $U=U_4.$ For the one-dimensional representations we have $U=1$.
\begin{table}%
\centering
\begin{minipage}{.4 \textwidth}%
\begin{tabular}{|c|c|c|c||c|c|}
\hline  & S & T&X& FS&Z(G)\\ 
\hline $\MoreRep{1}{1}$ & $1$ & $1$&$1$ &$1$  &$1$ \\ 
\hline $\MoreRep{1}{2}$ & $1$ & $\omega$&$1$ &$0$ &$1$ \\ 
\hline $\MoreRep{1}{3}$ & $1$ & $\omega^2$ &$1$&$0$  &$1$  \\ 
\hline
\hline $\MoreRep{4}{1}$ & $S_4$ & $T_4$ &  $X_4$&$1$ &$-1$ \\
\hline $\MoreRep{4}{2}$ & $S_4$ & $\omega^2 T_4$ &  $X_4$ &$0$ &$-1$\\ 
\hline $\MoreRep{4}{3}$ & $S_4$ & $\omega T_4$ &  $X_4$&$0$ &$-1$ \\   \hline
\end{tabular} 
\end{minipage}%
\qquad
\begin{minipage}{.4 \textwidth}%
\begin{tabular}{|c|c|c|c||c|c|}
\hline  & S & T&X& FS&Z(G)\\ 
\hline $\MoreRep{3}{1}$ & $S_3$ & $T_3$ &$\mathbbm{1}_3$&$1$ &$1$   \\ 
\hline $\MoreRep{3}{2}$ & $T_3 S_3 T_3^2$ & $T_3$ &$S_3$  &$1$ &$1$ \\ 
\hline $\MoreRep{3}{3}$ & $T_3 S_3 T_3^2$ & $T_3$ &$T_3^2 S_3 T_3$  &$1$  &$1$\\ 
\hline $\MoreRep{3}{4}$ & $\mathbbm{1}_3$ & $T_3$ &  $T_3 S_3 T_3^2$&$1$ &$1$ \\ 
\hline $\MoreRep{3}{5}$ & $T_3^2 S_3 T_3$ & $T_3$ &  $T_3^2 S_3 T_3$ &$1$ &$1$\\ 
  \hline
\end{tabular} 
\end{minipage}%
\caption{Representations of $Q_8\rtimes A_4$ in the chosen basis. The one-dimensional representations and the first three-dimensional one are the unfaithful $A_4$ representations  (therefore $X=\mathbbm{1}$), which the leptons are assigned to in Refs. \cite{Holthausen:2011vd,Holthausen:2012wz}. The representation $\MoreRep{4}{1}$ is used to break $A_4$ in the neutrino sector. Note that this representation is double valued, i.e. $Z(G)=X^2=-\mathbbm{1}$. FS is the Frobenius-Schur indicator $\frac{1}{\vert G \vert}\sum_{g \in G}\chi(g^2)$ that takes the values $1$ for real, $0$ for complex or $-1$ for pseudo-real representations, respectively. The matrices $S_3$ and $T_3$ have been defined in \Eqref{eq:Ma-basis}. \label{tab:Q8rtimesA4representations}}
\end{table}

Clearly the relation (\ref{eq:aut-def}) cannot be fulfilled by $\MoreRep{3}{1}$ as $\rho(X)=\mathbbm{1}_3$
\begin{equation}
1=U \rho(X) U^{-1}=\rho(SX)=S_3
\end{equation}
for any U. The representation $\MoreRep{3}{1}$ is rather part of a larger representation that also includes $\MoreRep{3}{5}$\footnote{For $\MoreRep{3}{5}$ we have $\rho(S)=\rho(X)$ and therefore \Eqref{eq:aut-def} would imply $\rho(S)=\rho(X)=\mathbbm{1}_3$.}:
\begin{equation}
S=\diag(S_3, T_3^2 S_3 T_3), \qquad T=\diag(T_3,T_3^2),\qquad X=\diag(\mathbbm{1}_3,T_3^2 S_3 T_3),\qquad
U=\left(
\begin{array}{cc}
0&T_3\\T_3^2&0
\end{array}
\right).
\end{equation}
The real representations $\MoreRep{3}{2,3,4}$ can be extended to representations of the CP-extended group by 
$
U=\tilde{U}_3. 
$
We have therefore seen that a CP transformation as defined in (\ref{eq:CP-def}) can only be realised if both $\MoreRep{3}{1}$ and $\MoreRep{3}{5}$ are present in the Lagrangian, i.e. the condition of CP conservation requires non-trivial relations among real representations of the group, something one would not immediately suspect. To summarise a consistent definition of CP acts as
\begin{equation}
\MoreRep{4}{i}\rightarrow U_4\MoreRep{4}{i}^*\qquad \MoreRep{3}{i}\rightarrow \tilde{U}_3\MoreRep{3}{f(i)}^*\qquad \MoreRep{1}{i}\rightarrow \MoreRep{1}{i}^*
\end{equation}
with $f:\{1,2,3,4,5\}\rightarrow \{5,2,3,4,1\}$.
\setcounter{nodecount}{0}
\begin{table}
\centering
\begin{tabular}{c|ccccccccccc}
&\tabnode{$E$}&\tabnode{$T$}& \tabnode{$SYX$} &\tabnode{$SY$}&\tabnode{$X^2$}&\tabnode{$T^2$}&\tabnode{$XT$}&\tabnode{$S$}&\tabnode{$SX$}&\tabnode{$X$}&\tabnode{$SXT^2$} \\ \hline
$\MoreRep{1}{1}$&  1&  1 & 1&  1 & 1 & 1 &  1 & 1  &1 & 1  & 1  \\
\tabnode{$\MoreRep{1}{2}$}&  1& $\omega$&  1&  1&  1&  $\omega^2$&  $\omega$&  1&  1&  1&   $\omega^2$ \\
\tabnode{$\MoreRep{1}{3}$}&  1&  $\omega^2$ & 1&  1 & 1 &$\omega$&   $\omega^2$&  1 & 1&  1 & $\omega$ \\
\tabnode{$\MoreRep{3}{1}$}&      3&  . &-1& -1&  3&  . &  . &-1 &-1 & 3  & . \\
\tabnode{$\MoreRep{3}{2}$}&      3&  . & 3 &-1 & 3 & .  & . &-1& -1 &-1  & . \\
\tabnode{$\MoreRep{3}{3}$}&      3&  . &-1&  3 & 3 & .  & . &-1& -1 &-1  & . \\
\tabnode{$\MoreRep{3}{4}$}&      3&  . &-1& -1&  3&  . &  .&  3& -1& -1  & . \\
\tabnode{$\MoreRep{3}{5}$}&      3&  . &-1& -1&  3&  . &  .& -1&  3& -1  & . \\
\tabnode{$\MoreRep{4}{1}$}&      4&  1&  . & . &-4&  1&  -1&  .&  .&  .&  -1 \\
\tabnode{$\MoreRep{4}{2}$}&     4 & $\omega^2$ & .  &. &-4 &$\omega$&  -$\omega^2$&  .&  .&  .& -$\omega$ \\
\tabnode{$\MoreRep{4}{3}$}&     4 &$\omega$&  . & . &-4&  $\omega^2$& -$\omega$&  .&  .&  .&  -$\omega^2$ \\
\end{tabular}
\caption{Character table of $Q_8\rtimes A_4$. The first line indicates
  representatives of the different conjugacy classes. Zeroes in the
  character table are denoted by a dot $.$  and $\omega$ is the third
  root of unity $\omega=e^{2\pi\I/3}$ and $Y=T^2 X T$. The arrows illustrate the generators of the outer automorphism group $h_4$(blue), $h_5$(red), $h_6$(green).\label{tab:ctblQ8A4}}
\begin{tikzpicture}[overlay]
\node [above=.3cm,minimum width=0pt] at (1) (eins){};
\node [above=.3cm,minimum width=0pt] at (2) (T){};
\node [above=.3cm,minimum width=0pt] at (3) (SYX){};
\node [above=.3cm,minimum width=0pt] at (4) (SY){};
\node [above=.3cm,minimum width=0pt] at (5) (XX){};
\node [above=.3cm,minimum width=0pt] at (6) (TT){};
\node [above=.3cm,minimum width=0pt] at (7) (XT){};
\node [above=.3cm,minimum width=0pt] at (8) (S){};
\node [above=.3cm,minimum width=0pt] at (9) (SX){};
\node [above=.3cm,minimum width=0pt] at (10) (X){};
\node [above=.3cm,minimum width=0pt] at (11) (SXTT){};
\node [left=.2cm,minimum width=0pt] at (12) (r12){};
\node [left=.2cm,minimum width=0pt] at (13) (r13){};
\node [left=.2cm,minimum width=0pt] at (14) (r31){};
\node [left=.2cm,minimum width=0pt] at (15) (r32){};
\node [left=.2cm,minimum width=0pt] at (16) (r33){};
\node [left=.2cm,minimum width=0pt] at (17) (r34){};
\node [left=.2cm,minimum width=0pt] at (18) (r35){};
\node [left=.2cm,minimum width=0pt] at (19) (r41){};
\node [left=.2cm,minimum width=0pt] at (20) (r42){};
\node [left=.2cm,minimum width=0pt] at (21) (r43){};
\draw [<->,out=45,in=135,blue!50, very thick,below=1cm] (T) to (TT);
\draw [<->,out=45,in=135,blue!50, very thick,below=1cm] (XT) to (SXTT);
\draw [<->,out=45,in=135,blue!50, very thick,below=1cm] (SX) to (X);
\draw [<->,out=225,in=135,blue!50, very thick,below=1cm] (r12) to (r13);
\draw [<->,out=225,in=135,blue!50, very thick,below=1cm] (r42) to (r43);
\draw [<->,out=225,in=135,blue!50, very thick,below=1cm] (r31) to (r35);
\draw [<->,out=60,in=120,red!50, very thick,below=1cm] (T) to (TT);
\draw [<->,out=60,in=120,red!50, very thick,below=1cm] (XT) to (SXTT);
\draw [<->,out=60,in=120,red!50, very thick,below=1cm] (SYX) to (SY);
\draw [<->,out=180,in=180,red!50, very thick,below=1cm] (r12) to (r13);
\draw [<->,out=180,in=180,red!50, very thick,below=1cm] (r42) to (r43);
\draw [<->,out=225,in=135,red!50, very thick,below=1cm] (r32) to (r33);
\draw [->,out=45,in=135,green!50, very thick,below=1cm] (SYX) to (S);
\draw [<-,out=30,in=150,green!50, very thick,below=1cm]  (SYX) to (SY) ;
\draw [<-,out=30,in=150,green!50, very thick,below=1cm]  (SY) to (S) ;
\draw [<-,out=225,in=135,green!50, very thick,below=1cm] (r33) to (r34);
\draw [<-,out=180,in=180,green!50, very thick,below=1cm] (r32) to (r33);
\draw [->,out=225,in=135,green!50, very thick,below=1cm,tension=.1] (r32) to (r34);
\end{tikzpicture}
\end{table}

The natural question is now if it is possible to have outer automorphisms of the group that act as CP in the sense that they interchange the complex representations $\MoreRep{1}{2,3}$ and $\MoreRep{4}{2,3}$ but transform the real representations only within themselves. This question can be answered using the explicit form of the generators of \Eqref{eq:Q8aut-gen}.

An outer automorphism swaps conjugacy classes and representations in such a way as to leave the character table \ref{tab:ctblQ8A4} invariant. For illustration look at the automorphism $h_4$ (\ref{eq:aut-def}). It acts on the conjugacy classes as 
\begin{equation}
G\cdot T \leftrightarrow G\cdot T^2,\qquad G\cdot XT \leftrightarrow G  \cdot  SXT^2,\qquad G \cdot X \leftrightarrow G \cdot SX 
\end{equation}
where $G\cdot T\equiv \{ g T g^{-1}:g \in G\}$, leaving all other conjugacy classes invariant. To obtain a symmetry of the character table one therefore needs to interchange the representations
\begin{equation}
\MoreRep{1}{2}\leftrightarrow  \MoreRep{1}{3},\qquad \MoreRep{4}{2}\leftrightarrow  \MoreRep{4}{3},\qquad\MoreRep{3}{1}\leftrightarrow  \MoreRep{3}{5}.
\end{equation}
If we want to have a symmetry of the character table without interchanging any real representations that still acts as CP, we therefore have to have an automorphism that realises 
\begin{equation}
G\cdot T \leftrightarrow G\cdot T^2,\qquad G\cdot XT \leftrightarrow G\cdot SXT^2
\end{equation}
while keeping all other conjugacy classes invariant. No such automorphism exists, as can be inferred from \Eqref{eq:Q8aut-gen}.\footnote{It is convenient to use the computer algebra system \GAP \cite{GAP4}.}

However, if we relax the condition to the point where we only demand that the representation $\MoreRep{3}{1}$ transforms into itself we have to search for outer automorphisms that realise
\begin{equation}
G\cdot T \leftrightarrow G\cdot T^2,\qquad G\cdot XT\leftrightarrow G\cdot SXT^2\qquad G\cdot X \leftrightarrow G\cdot X.
\end{equation}
Indeed there is a automorphism that realises this:
$
h_5:(S,T,X)\to(S,T^2,X^3)\;.
$
An explicit matrix representation for representation $\MoreRep{4}{1}$ is given by
\begin{align}
U_4(h_5)=\frac{1}{2}\left(
\begin{array}{cccc}
 1 & -1 & 1 & -1 \\
 -1 & 1 & 1 & -1 \\
 1 & 1 & -1 & -1 \\
 -1 & -1 & -1 & -1
\end{array}
\right)
\end{align}
and for the representation $\MoreRep{3}{1} $ we find $U=U_3$.

Having found a consistent CP transformation for a theory that contains only the representations $\MoreRep{3}{1}$, $\MoreRep{4}{1}$ and $\MoreRep{1}{i}$ we can now ask ourselves the question that lead us to this study of generalised CP transformations. Namely if we take the flavon content of Ref.~\cite{Holthausen:2011vd}  and promote all the fields to electroweak (EW) doublets $\chi\sim\MoreRep{3}{1}$ and $\phi\sim\MoreRep{4}{1}$, there is a purely imaginary coupling\footnote{$\left( \chi^\dagger \chi  \right)_{\MoreRep{3}{1,S}}$ is real and $\left( \phi^\dagger \phi \right)_{\MoreRep{3}{1}}$ is purely imaginary, therefore the coupling has to be purely imaginary.} 
\begin{align}
\lambda \left( \chi^\dagger \chi  \right)_{\MoreRep{3}{1,S}}\cdot \left( \phi^\dagger \phi \right)_{\MoreRep{3}{1}}+\hc.
\label{eq:dangerous-term-motivation}
\end{align}
which breaks the accidental symmetry needed for vacuum alignment~\cite{Holthausen:2011vd}. To forbid this imaginary coupling one might think that a CP symmetry can be invoked. However, the consistent CP transformation corresponding to $h_5$ (which is unique up to inner automorphisms) under which the EW doublets $\phi$ and $\chi$ transform as
\begin{align}
\phi_{i}\rightarrow U_4 \phi^*, \qquad \chi \rightarrow U_3 \chi^*
\end{align}
leaves the operator in \Eqref{eq:dangerous-term-motivation}  invariant, even though it is purely imaginary. 

For completeness we also give a representation of $h_6$
\begin{align}
U_4(h_6)=\frac{1}{2}\left(
\begin{array}{cccc}
 1 & -1 & -1 & -1 \\
 -1 & 1 & -1 & -1 \\
 -1 & -1 & 1 & -1 \\
 1 & 1 & 1 & -1
\end{array}
\right)
\end{align}
from which all the other representation matrices can be derived using the Clebsch-Gordon coefficients.


\mathversion{bold}
\subsection{$S_4\cong (Z_2\times Z_2)\rtimes S_3\cong\SG{24}{12}$}
\mathversion{normal}
There is a complete classification of automorphism groups for the symmetric group $S_n$, which we summarise in \Tabref{tab:Sn}. Since $\out{S_n}=Z_1$ for $n\neq 2,6$, there is no non-trivial generalised CP transformation, especially $S_3$ and $S_4$, which have been introduced in~\cite{Pakvasa:1977in,Pakvasa:1978tx} and used in models explaining the leptonic mixing structure, do not allow for a non-trivial generalised CP transformation. The recently discussed generalised CP transformation in Ref.~\cite{Feruglio:2012cw} is an inner automorphism of $S_4$, similarly the generalised CP in the framework of $S_3$ discussed in~\cite{Adler:1998as}. Obviously, it is always possible to apply a group transformation at the same time as a CP transformation.  We therefore do not discuss $S_3$ or $S_4$ in more detail.
\begin{table}[tb]\centering
\begin{tabular}{l|cccc}
& $\zentrum{S_n}$ & $\aut{S_n}$ & $\inn{S_n}$ & $\out{S_n}$ \\\hline
$n\neq 2,6$ & $Z_1$ & $S_n$ & $S_n$ & $Z_1$\\
$n=2$ & $Z_2$ & $Z_1$ & $Z_1$ & $Z_1$\\
$n=6$ & $Z_1$ & $S_6\rtimes Z_2$ & $S_6$ & $Z_2$\\
\end{tabular}
\caption{Group structure of the symmetric group $S_n$\label{tab:Sn}}
\end{table}

\mathversion{bold}
\subsection{$T_7\cong Z_7\rtimes Z_3\cong\SG{21}{1}$}
\mathversion{normal}
The group $T_7\cong Z_7\rtimes Z_3\cong\SG{21}{1}=\braket{A,B|A^7=B^3=BAB^{-1}A^5=E}$ has been first used in particle physics in Ref.~\cite{Luhn:2007sy}. 
In the basis used in~\cite{Hagedorn:2008bc}, the generators $A$ and $B$ are given by 
\begin{equation}
\rho(A)=\diag(\eta,\,\eta^2,\,\eta^4)
\qquad
\rho(B)=T_3
\end{equation}
for \MoreRep{3}{1} with $\eta=e^{2\pi\I/7}$. $T_7$ has a trivial centre and therefore the inner automorphism group $\inn{T_7}$ is isomorphic to $T_7$ itself. However, since $\rho(A)^*=\rho(A^6)\in \im \rho$ and  $\rho(B)^*=\rho(B)\in \im \rho$, the outer automorphism group is non-trivial. Its generator $u:(A,B)\rightarrow (A^6,B)$ is thus represented by the identity matrix on the three dimensional representation and this basis is thus a CP basis. Concluding the structure of the automorphism group is described by
\begin{align}
\zentrum{T_7}&\cong Z_1&
\aut{T_7}&\cong \SG{42}{2}\\\nonumber
\inn{T_7}&\cong T_7&
\out{T_7}&\cong Z_2\;.
\end{align}
The outer automorphism exchanges the three-dimensional representations, while leaving the one-dimensional ones fixed, i.e.
\begin{equation}
\MoreRep{1}{2}\rightarrow \MoreRep{1}{2},\qquad 
\MoreRep{1}{3}\rightarrow \MoreRep{1}{3}\qquad \mathrm{and}\qquad
\Rep{3}\leftrightarrow \Rep{3}^*\;.
\end{equation}

\mathversion{bold}
\subsection{$\Delta(108)\cong\SG{108}{22}$ (or $\Delta(216)\cong \SG{216}{95}$)}
\mathversion{normal}

Recently~\cite{Ferreira:2012ri}, CP violation has been discussed in the context of $\Delta(108)=\Delta(3\times 6^2)$~\footnote{$\Delta(108)$ has been first used in the lepton sector in Ref.~\cite{deMedeirosVarzielas:2005qg}. There is a comprehensive study of $\Delta(3n^2)$\cite{Luhn:2007ul} and $\Delta(6n^2)$\cite{Escobar:2008vc} groups in the context of flavour symmetries.}, which may be represented by a faithful three-dimensional representation as 
\begin{align}
\rho(\mathcal{S})=S_3, \qquad \rho(\mathcal{T})=T_3 \qquad \rho(\mathcal{T}')=\diag(1, \omega , \omega^2)\;.
\end{align}
The model possesses an accidental $\mu-\tau$ exchange symmetry, which is generated by $U_3$~\footnote{The matrices $S_3$, $T_3$ and $U_3$ have been defined in \Eqref{eq:Ma-basis} and \Eqref{eq:defU3}.}. Including this generator $U=U_3$, the group becomes $\Delta(6\times 6^2)$. A generalised CP transformation was defined on the faithful representation $\ell_R$ as 
\begin{equation}
\ell_R \rightarrow i U_3 \ell_R^*,
\end{equation}
where we have suppressed the Lorentz structure. This is equivalent to the automorphism $u:(\mathcal{S},\mathcal{T},\mathcal{T}')\rightarrow (\mathcal{S},\mathcal{T}^2,\mathcal{T}')$, which is outer in $\Delta(3\times 6^2)$ and inner in $\Delta(6\times 6^2)$. In Ref.~\cite{Ferreira:2012ri} this has been consistently applied to all non-faithful representations which they consider.

Let us comment on the origin of maximal CP violation in their model, which seems to be in conflict with our general statement that there can be no CP violation. It is related to the breaking of the flavour symmetry in their model. One of the scalar fields breaking the flavour symmetry is the scalar $\phi$ transforming as 
\begin{align}
\rho(\mathcal{S})=S_3, \qquad \rho(\mathcal{T})=T_3 \qquad \rho(\mathcal{T}')=\mathbbm{1}_3\;,
\end{align}
and thus transforms only under the subgroup $\vev{\mathcal{S},\mathcal{T}}\cong A_4$ with the CP transformation $\phi\rightarrow U_3 \phi^*$. CP conservation would therefore require $v_2=v_3^*$. However, they have to assume a large hierarchy in the VEVs of $\phi$ in order to accommodate the hierarchy in the charged lepton sector, which is given by
$m_e:m_\mu:m_\tau = v_1:v_2:v_3$. Hence, the requirement $\abs{v_2}/\abs{v_3}=m_\mu/m_\tau\ll 1$ is the necessary ingredient for maximal CP violation in the model. 

\section{Conclusions and Outlook\label{sec:conclusions}}
We have given consistency conditions for the definition of CP in theories with discrete flavour symmetries that have sometimes been overlooked in the literature. We have shown that every generalised CP transformation furnishes a representation of an outer automorphism and that generalised CP invariance implies vanishing CP phases. We have applied these ideas to popular flavour groups with three-dimensional representations and group order smaller than 31.\footnote{For completeness, we mention the group $A_4\times Z_2\cong\SG{24}{13}$, which we did not discuss in detail. It has been mentioned in the survey of Ref.~\cite{Parattu:2010cy}. Its automorphism group structure is directly inherited from $A_4$ with the addition that it has a non-trivial centre $\zentrum{Z_2}$.} In particular, we have shown that there is one unique non-trivial CP transformation (up to group transformations) for the group $T^\prime$,  which we applied to the models discussed in Ref.~\cite{Chen:2009lr,Meroni:2012ty}. We show that this CP is spontaneously broken by the VEVs of the doublets.
The claimed geometric CP-violation in Ref.~\cite{Chen:2009lr} can only be viewed as an arbitrary basis-dependent explicit breaking of CP.
In the case of $\Delta(27)$ we have shown that the so-called geometric phases may be viewed as the result of an accidental generalised CP transformation of the scalar potential. 
Finally, we showed in the case of $A_4$ that the phase of $(\chi^\dagger\chi)_{\MoreRep{3}{1}}(\chi^\dagger\chi)_{\MoreRep{3}{1}}$ in the potential of a single triplet does not break CP, which has also been independently shown in Ref.~\cite{Adelhart-Toorop:2012PhD,Ivanov:2012fp,Degee:2012sk}.
This clarifies the recent observation that CP conserving solutions result from seemingly explicitly CP-breaking potentials~\cite{Adelhart-Toorop:2010fkv3}.

The (outer) automorphism structure of small groups is very rich and it stands to wonder if not more physics might be hidden in there. We may speculate about this possibility in the following. $S_4$ is the smallest group that can really generate TBM (with all the caveats involved) and it is isomorphic to the automorphism group of $A_4$. Maybe the accidental symmetry that makes $A_4$ to $S_4$ on the level of mass matrices is connected
to this fact. This would open an interesting avenue for model building: interesting mixing patterns can be obtained from $\Delta(6 n^2)$ but it is quite unappealing to start from such large groups, it might be nicer to start from smaller groups and obtain the accidental symmetry from the larger automorphism group in the same way as in $A_4$ models. As an example how complicated structures can arise from simpler ones, look at the automorphism group of $\Delta(27)$, which is of order $432$. The smallest group whose automorphism group contains $\Delta(96)$ is given by $(Z_4\times Z_4)\rtimes Z_2\cong \SG{32}{34}$. Further investigation of these ideas is left for future work.

\section*{Note added}
While this work was being finalised, a related work~\cite{Feruglio:2012cw} addressing CP in the context of discrete flavour symmetries appeared on the arxiv. We both give a general definition and discussion of generalised CP symmetries. Our work differs from~\cite{Feruglio:2012cw} as follows. They consider the physical implications for the lepton mixing parameters of a remnant CP symmetry in the neutrino sector. In particular, they discuss the groups $S_4$ as well as $A_4$ in more detail. We, on the other hand, emphasise the relation of generalised CP transformations to the automorphism group and especially the outer automorphism group. We perform a systematic study of all generalised CP transformations for all groups with a three-dimensional irreducible representation of order less than 31.
In particular, we discuss the "calculable phases" in models based on $\Delta(27)$ and interpret them in terms of an accidental generalised CP transformation as well as comment on the recent claims of geometric CP violation in the context of $T^\prime$ models.

\section*{Acknowledgements}
We would like to thank I.~de Medeiros Varzielas for comments on the manuscript. Furthermore, we thank C.~Hagedorn, A.~Meroni, S.~Petcov and especially M.~Spinrath for discussions on CP violation in $T^\prime$ models. We thank the referee for his detailed comments. We thank M.~Ratz and A.~Trautner for pointing out some typos and stimulating discussions.
M.S.  would like to acknowledge MPI f\"ur Kernphysik, where a part of this work has been done, for hospitality of its staff and the generous
support. M.H.~acknowledges support by the International Max Planck
Research School for Precision Tests of Fundamental Symmetries. This
work was supported in part by the Australian Research Council.

\bibliography{CP}

\begin{thebibliography}{64}
\expandafter\ifx\csname natexlab\endcsname\relax\def\natexlab#1{#1}\fi
\expandafter\ifx\csname bibnamefont\endcsname\relax
  \def\bibnamefont#1{#1}\fi
\expandafter\ifx\csname bibfnamefont\endcsname\relax
  \def\bibfnamefont#1{#1}\fi
\expandafter\ifx\csname citenamefont\endcsname\relax
  \def\citenamefont#1{#1}\fi
\expandafter\ifx\csname url\endcsname\relax
  \def\url#1{\texttt{#1}}\fi
\expandafter\ifx\csname urlprefix\endcsname\relax\def\urlprefix{URL }\fi
\providecommand{\bibinfo}[2]{#2}
\providecommand{\eprint}[2][]{\url{#2}}

\bibitem[{\citenamefont{Abe et~al.}(2012)}]{Abe:2011fz}
\bibinfo{author}{\bibfnamefont{Y.}~\bibnamefont{Abe}} \bibnamefont{et~al.}
  (\bibinfo{collaboration}{DOUBLE-CHOOZ Collaboration}),
  \bibinfo{journal}{Phys.Rev.Lett.} \textbf{\bibinfo{volume}{108}},
  \bibinfo{pages}{131801} (\bibinfo{year}{2012}), \eprint{1112.6353}.

\bibitem[{\citenamefont{An et~al.}(2012)}]{An:2012eh}
\bibinfo{author}{\bibfnamefont{F.}~\bibnamefont{An}} \bibnamefont{et~al.}
  (\bibinfo{collaboration}{DAYA-BAY Collaboration}),
  \bibinfo{journal}{Phys.Rev.Lett.} \textbf{\bibinfo{volume}{108}},
  \bibinfo{pages}{171803} (\bibinfo{year}{2012}), \eprint{1203.1669}.

\bibitem[{\citenamefont{Ahn et~al.}(2012)}]{Ahn:2012nd}
\bibinfo{author}{\bibfnamefont{J.}~\bibnamefont{Ahn}} \bibnamefont{et~al.}
  (\bibinfo{collaboration}{RENO collaboration}),
  \bibinfo{journal}{Phys.Rev.Lett.} \textbf{\bibinfo{volume}{108}},
  \bibinfo{pages}{191802} (\bibinfo{year}{2012}), \eprint{1204.0626}.

\bibitem[{\citenamefont{Branco et~al.}(1984)\citenamefont{Branco, Gerard, and
  Grimus}}]{Branco:1983tn}
\bibinfo{author}{\bibfnamefont{G.}~\bibnamefont{Branco}},
  \bibinfo{author}{\bibfnamefont{J.}~\bibnamefont{Gerard}}, \bibnamefont{and}
  \bibinfo{author}{\bibfnamefont{W.}~\bibnamefont{Grimus}},
  \bibinfo{journal}{Phys.Lett.} \textbf{\bibinfo{volume}{B136}},
  \bibinfo{pages}{383} (\bibinfo{year}{1984}).

\bibitem[{\citenamefont{de~Medeiros~Varzielas and
  Emmanuel-Costa}(2011)}]{deMedeirosVarzielas:2011zw}
\bibinfo{author}{\bibfnamefont{I.}~\bibnamefont{de~Medeiros~Varzielas}}
  \bibnamefont{and}
  \bibinfo{author}{\bibfnamefont{D.}~\bibnamefont{Emmanuel-Costa}},
  \bibinfo{journal}{Phys.Rev.} \textbf{\bibinfo{volume}{D84}},
  \bibinfo{pages}{117901} (\bibinfo{year}{2011}), \eprint{1106.5477}.

\bibitem[{\citenamefont{de~Medeiros~Varzielas
  et~al.}(2012)\citenamefont{de~Medeiros~Varzielas, Emmanuel-Costa, and
  Leser}}]{de-Medeiros-Varzielas:2012fk}
\bibinfo{author}{\bibfnamefont{I.}~\bibnamefont{de~Medeiros~Varzielas}},
  \bibinfo{author}{\bibfnamefont{D.}~\bibnamefont{Emmanuel-Costa}},
  \bibnamefont{and} \bibinfo{author}{\bibfnamefont{P.}~\bibnamefont{Leser}},
  \bibinfo{journal}{Phys.Lett.} \textbf{\bibinfo{volume}{B716}},
  \bibinfo{pages}{193} (\bibinfo{year}{2012}), \eprint{1204.3633}.

\bibitem[{\citenamefont{de~Medeiros~Varzielas}(2012)}]{Varzielas:2012pd}
\bibinfo{author}{\bibfnamefont{I.}~\bibnamefont{de~Medeiros~Varzielas}},
  \bibinfo{journal}{JHEP} \textbf{\bibinfo{volume}{1208}}, \bibinfo{pages}{055}
  (\bibinfo{year}{2012}), \eprint{1205.3780}.

\bibitem[{\citenamefont{Bhattacharyya et~al.}(2012)\citenamefont{Bhattacharyya,
  de~Medeiros~Varzielas, and Leser}}]{Bhattacharyya:2012pi}
\bibinfo{author}{\bibfnamefont{G.}~\bibnamefont{Bhattacharyya}},
  \bibinfo{author}{\bibfnamefont{I.}~\bibnamefont{de~Medeiros~Varzielas}},
  \bibnamefont{and} \bibinfo{author}{\bibfnamefont{P.}~\bibnamefont{Leser}},
  \bibinfo{journal}{Phys.Rev.Lett.} \textbf{\bibinfo{volume}{109}},
  \bibinfo{pages}{241603} (\bibinfo{year}{2012}), \eprint{1210.0545}.

\bibitem[{\citenamefont{Ferreira et~al.}(2012)\citenamefont{Ferreira, Grimus,
  Lavoura, and Ludl}}]{Ferreira:2012ri}
\bibinfo{author}{\bibfnamefont{P.}~\bibnamefont{Ferreira}},
  \bibinfo{author}{\bibfnamefont{W.}~\bibnamefont{Grimus}},
  \bibinfo{author}{\bibfnamefont{L.}~\bibnamefont{Lavoura}}, \bibnamefont{and}
  \bibinfo{author}{\bibfnamefont{P.}~\bibnamefont{Ludl}},
  \bibinfo{journal}{JHEP} \textbf{\bibinfo{volume}{1209}}, \bibinfo{pages}{128}
  (\bibinfo{year}{2012}), \eprint{1206.7072}.

\bibitem[{\citenamefont{Chen and Mahanthappa}(2009)}]{Chen:2009lr}
\bibinfo{author}{\bibfnamefont{M.-C.} \bibnamefont{Chen}} \bibnamefont{and}
  \bibinfo{author}{\bibfnamefont{K.}~\bibnamefont{Mahanthappa}},
  \bibinfo{journal}{Phys.Lett.} \textbf{\bibinfo{volume}{B681}},
  \bibinfo{pages}{444} (\bibinfo{year}{2009}), \eprint{0904.1721}.

\bibitem[{\citenamefont{Meroni et~al.}(2012)\citenamefont{Meroni, Petcov, and
  Spinrath}}]{Meroni:2012ty}
\bibinfo{author}{\bibfnamefont{A.}~\bibnamefont{Meroni}},
  \bibinfo{author}{\bibfnamefont{S.}~\bibnamefont{Petcov}}, \bibnamefont{and}
  \bibinfo{author}{\bibfnamefont{M.}~\bibnamefont{Spinrath}},
  \bibinfo{journal}{Phys.Rev.} \textbf{\bibinfo{volume}{D86}},
  \bibinfo{pages}{113003} (\bibinfo{year}{2012}), \eprint{1205.5241}.

\bibitem[{\citenamefont{Lee}(1973)}]{Lee:1973iz}
\bibinfo{author}{\bibfnamefont{T.}~\bibnamefont{Lee}},
  \bibinfo{journal}{Phys.Rev.} \textbf{\bibinfo{volume}{D8}},
  \bibinfo{pages}{1226} (\bibinfo{year}{1973}).

\bibitem[{\citenamefont{Branco}(1980)}]{Branco:1979pv}
\bibinfo{author}{\bibfnamefont{G.~C.} \bibnamefont{Branco}},
  \bibinfo{journal}{Phys.Rev.Lett.} \textbf{\bibinfo{volume}{44}},
  \bibinfo{pages}{504} (\bibinfo{year}{1980}).

\bibitem[{\citenamefont{Haber and Surujon}(2012)}]{Haber:2012np}
\bibinfo{author}{\bibfnamefont{H.~E.} \bibnamefont{Haber}} \bibnamefont{and}
  \bibinfo{author}{\bibfnamefont{Z.}~\bibnamefont{Surujon}},
  \bibinfo{journal}{Phys.Rev.} \textbf{\bibinfo{volume}{D86}},
  \bibinfo{pages}{075007} (\bibinfo{year}{2012}), \eprint{1201.1730}.

\bibitem[{\citenamefont{Ecker et~al.}(1981)\citenamefont{Ecker, Grimus, and
  Konetschny}}]{Ecker:1981wv}
\bibinfo{author}{\bibfnamefont{G.}~\bibnamefont{Ecker}},
  \bibinfo{author}{\bibfnamefont{W.}~\bibnamefont{Grimus}}, \bibnamefont{and}
  \bibinfo{author}{\bibfnamefont{W.}~\bibnamefont{Konetschny}},
  \bibinfo{journal}{Nucl.Phys.} \textbf{\bibinfo{volume}{B191}},
  \bibinfo{pages}{465} (\bibinfo{year}{1981}).

\bibitem[{\citenamefont{Ecker et~al.}(1984)\citenamefont{Ecker, Grimus, and
  Neufeld}}]{Ecker:1983hz}
\bibinfo{author}{\bibfnamefont{G.}~\bibnamefont{Ecker}},
  \bibinfo{author}{\bibfnamefont{W.}~\bibnamefont{Grimus}}, \bibnamefont{and}
  \bibinfo{author}{\bibfnamefont{H.}~\bibnamefont{Neufeld}},
  \bibinfo{journal}{Nucl.Phys.} \textbf{\bibinfo{volume}{B247}},
  \bibinfo{pages}{70} (\bibinfo{year}{1984}).

\bibitem[{\citenamefont{Neufeld et~al.}(1988)\citenamefont{Neufeld, Grimus, and
  Ecker}}]{Neufeld:1987wa}
\bibinfo{author}{\bibfnamefont{H.}~\bibnamefont{Neufeld}},
  \bibinfo{author}{\bibfnamefont{W.}~\bibnamefont{Grimus}}, \bibnamefont{and}
  \bibinfo{author}{\bibfnamefont{G.}~\bibnamefont{Ecker}},
  \bibinfo{journal}{Int.J.Mod.Phys.} \textbf{\bibinfo{volume}{A3}},
  \bibinfo{pages}{603} (\bibinfo{year}{1988}).

\bibitem[{\citenamefont{Grimus and Rebelo}(1997)}]{Grimus:1997fk}
\bibinfo{author}{\bibfnamefont{W.}~\bibnamefont{Grimus}} \bibnamefont{and}
  \bibinfo{author}{\bibfnamefont{M.}~\bibnamefont{Rebelo}},
  \bibinfo{journal}{Phys.Rept.} \textbf{\bibinfo{volume}{281}},
  \bibinfo{pages}{239} (\bibinfo{year}{1997}), \eprint{hep-ph/9506272}.

\bibitem[{\citenamefont{Harrison and
  Scott}(2002{\natexlab{a}})}]{Harrison:2002et}
\bibinfo{author}{\bibfnamefont{P.}~\bibnamefont{Harrison}} \bibnamefont{and}
  \bibinfo{author}{\bibfnamefont{W.}~\bibnamefont{Scott}},
  \bibinfo{journal}{Phys.Lett.} \textbf{\bibinfo{volume}{B547}},
  \bibinfo{pages}{219} (\bibinfo{year}{2002}{\natexlab{a}}),
  \eprint{hep-ph/0210197}.

\bibitem[{\citenamefont{Harrison and
  Scott}(2002{\natexlab{b}})}]{Harrison:2002kp}
\bibinfo{author}{\bibfnamefont{P.}~\bibnamefont{Harrison}} \bibnamefont{and}
  \bibinfo{author}{\bibfnamefont{W.}~\bibnamefont{Scott}},
  \bibinfo{journal}{Phys.Lett.} \textbf{\bibinfo{volume}{B535}},
  \bibinfo{pages}{163} (\bibinfo{year}{2002}{\natexlab{b}}),
  \eprint{hep-ph/0203209}.

\bibitem[{\citenamefont{Harrison and Scott}(2004)}]{Harrison:2004he}
\bibinfo{author}{\bibfnamefont{P.}~\bibnamefont{Harrison}} \bibnamefont{and}
  \bibinfo{author}{\bibfnamefont{W.}~\bibnamefont{Scott}},
  \bibinfo{journal}{Phys.Lett.} \textbf{\bibinfo{volume}{B594}},
  \bibinfo{pages}{324} (\bibinfo{year}{2004}), \eprint{hep-ph/0403278}.

\bibitem[{\citenamefont{Grimus and Lavoura}(2004)}]{Grimus:2003yn}
\bibinfo{author}{\bibfnamefont{W.}~\bibnamefont{Grimus}} \bibnamefont{and}
  \bibinfo{author}{\bibfnamefont{L.}~\bibnamefont{Lavoura}},
  \bibinfo{journal}{Phys.Lett.} \textbf{\bibinfo{volume}{B579}},
  \bibinfo{pages}{113} (\bibinfo{year}{2004}), \eprint{hep-ph/0305309}.

\bibitem[{\citenamefont{Farzan and Smirnov}(2007)}]{Farzan:2006vj}
\bibinfo{author}{\bibfnamefont{Y.}~\bibnamefont{Farzan}} \bibnamefont{and}
  \bibinfo{author}{\bibfnamefont{A.~Y.} \bibnamefont{Smirnov}},
  \bibinfo{journal}{JHEP} \textbf{\bibinfo{volume}{0701}}, \bibinfo{pages}{059}
  (\bibinfo{year}{2007}), \eprint{hep-ph/0610337}.

\bibitem[{\citenamefont{Joshipura et~al.}(2009)\citenamefont{Joshipura,
  Kodrani, and Patel}}]{Joshipura:2009tg}
\bibinfo{author}{\bibfnamefont{A.~S.} \bibnamefont{Joshipura}},
  \bibinfo{author}{\bibfnamefont{B.~P.} \bibnamefont{Kodrani}},
  \bibnamefont{and} \bibinfo{author}{\bibfnamefont{K.~M.} \bibnamefont{Patel}},
  \bibinfo{journal}{Phys.Rev.} \textbf{\bibinfo{volume}{D79}},
  \bibinfo{pages}{115017} (\bibinfo{year}{2009}), \eprint{0903.2161}.

\bibitem[{\citenamefont{Grimus and Lavoura}(2012)}]{Grimus:2012hu}
\bibinfo{author}{\bibfnamefont{W.}~\bibnamefont{Grimus}} \bibnamefont{and}
  \bibinfo{author}{\bibfnamefont{L.}~\bibnamefont{Lavoura}}
  (\bibinfo{year}{2012}), \eprint{1207.1678}.

\bibitem[{\citenamefont{Mohapatra and Nishi}(2012)}]{Mohapatra:2012tb}
\bibinfo{author}{\bibfnamefont{R.}~\bibnamefont{Mohapatra}} \bibnamefont{and}
  \bibinfo{author}{\bibfnamefont{C.}~\bibnamefont{Nishi}},
  \bibinfo{journal}{Phys.Rev.} \textbf{\bibinfo{volume}{D86}},
  \bibinfo{pages}{073007} (\bibinfo{year}{2012}), \eprint{1208.2875}.

\bibitem[{\citenamefont{Krishnan et~al.}(2012)\citenamefont{Krishnan, Harrison,
  and Scott}}]{Krishnan:2012me}
\bibinfo{author}{\bibfnamefont{R.}~\bibnamefont{Krishnan}},
  \bibinfo{author}{\bibfnamefont{P.}~\bibnamefont{Harrison}}, \bibnamefont{and}
  \bibinfo{author}{\bibfnamefont{W.}~\bibnamefont{Scott}}
  (\bibinfo{year}{2012}), \eprint{1211.2000}.

\bibitem[{\citenamefont{Feruglio et~al.}(2012)\citenamefont{Feruglio, Hagedorn,
  and Ziegler}}]{Feruglio:2012cw}
\bibinfo{author}{\bibfnamefont{F.}~\bibnamefont{Feruglio}},
  \bibinfo{author}{\bibfnamefont{C.}~\bibnamefont{Hagedorn}}, \bibnamefont{and}
  \bibinfo{author}{\bibfnamefont{R.}~\bibnamefont{Ziegler}}
  (\bibinfo{year}{2012}), \eprint{1211.5560}.

\bibitem[{\citenamefont{Babu and Kubo}(2005)}]{Babu:2004tn}
\bibinfo{author}{\bibfnamefont{K.~S.} \bibnamefont{Babu}} \bibnamefont{and}
  \bibinfo{author}{\bibfnamefont{J.}~\bibnamefont{Kubo}},
  \bibinfo{journal}{Phys.Rev.} \textbf{\bibinfo{volume}{D71}},
  \bibinfo{pages}{056006} (\bibinfo{year}{2005}), \eprint{hep-ph/0411226}.

\bibitem[{\citenamefont{Babu et~al.}(2011)\citenamefont{Babu, Kawashima, and
  Kubo}}]{Babu:2011mv}
\bibinfo{author}{\bibfnamefont{K.}~\bibnamefont{Babu}},
  \bibinfo{author}{\bibfnamefont{K.}~\bibnamefont{Kawashima}},
  \bibnamefont{and} \bibinfo{author}{\bibfnamefont{J.}~\bibnamefont{Kubo}},
  \bibinfo{journal}{Phys.Rev.} \textbf{\bibinfo{volume}{D83}},
  \bibinfo{pages}{095008} (\bibinfo{year}{2011}), \eprint{1103.1664}.

\bibitem[{GAP()}]{GAP4}
GAP, \emph{\bibinfo{title}{{GAP -- Groups, Algorithms, and Programming, Version
  4.5.5}}}, \bibinfo{organization}{The GAP~Group} (\bibinfo{year}{2012}),
  \urlprefix\url{http://www.gap-system.org)}.

\bibitem[{\citenamefont{H.U.Besche et~al.}(2002)\citenamefont{H.U.Besche,
  B.Eick, and E.O'Brien}}]{SmallGroups:2011}
\bibinfo{author}{\bibnamefont{H.U.Besche}},
  \bibinfo{author}{\bibnamefont{B.Eick}}, \bibnamefont{and}
  \bibinfo{author}{\bibnamefont{E.O'Brien}},
  \emph{\bibinfo{title}{{SmallGroups} - library of all 'small' groups, {GAP}
  package, Version included in {GAP} 4.5.5}}, \bibinfo{organization}{The
  GAP~Group} (\bibinfo{year}{2002}),
  \urlprefix\url{http://www.gap-system.org/Packages/sgl.html}.

\bibitem[{\citenamefont{Fairbairn and Fulton}(1982)}]{Fairbairn:1982jx}
\bibinfo{author}{\bibfnamefont{W.}~\bibnamefont{Fairbairn}} \bibnamefont{and}
  \bibinfo{author}{\bibfnamefont{T.}~\bibnamefont{Fulton}},
  \bibinfo{journal}{J.Math.Phys.} \textbf{\bibinfo{volume}{23}},
  \bibinfo{pages}{1747} (\bibinfo{year}{1982}).

\bibitem[{\citenamefont{Grimus and Ludl}(2012)}]{Grimus:2011ff}
\bibinfo{author}{\bibfnamefont{W.}~\bibnamefont{Grimus}} \bibnamefont{and}
  \bibinfo{author}{\bibfnamefont{P.~O.} \bibnamefont{Ludl}},
  \bibinfo{journal}{J.Phys.} \textbf{\bibinfo{volume}{A45}},
  \bibinfo{pages}{233001} (\bibinfo{year}{2012}), \eprint{1110.6376}.

\bibitem[{\citenamefont{Ishimori et~al.}(2012)\citenamefont{Ishimori,
  Kobayashi, Ohki, Okada, Shimizu et~al.}}]{Ishimori:2012zz}
\bibinfo{author}{\bibfnamefont{H.}~\bibnamefont{Ishimori}},
  \bibinfo{author}{\bibfnamefont{T.}~\bibnamefont{Kobayashi}},
  \bibinfo{author}{\bibfnamefont{H.}~\bibnamefont{Ohki}},
  \bibinfo{author}{\bibfnamefont{H.}~\bibnamefont{Okada}},
  \bibinfo{author}{\bibfnamefont{Y.}~\bibnamefont{Shimizu}},
  \bibnamefont{et~al.}, \bibinfo{journal}{Lect.Notes Phys.}
  \textbf{\bibinfo{volume}{858}}, \bibinfo{pages}{1} (\bibinfo{year}{2012}).

\bibitem[{\citenamefont{Branco et~al.}(2012)\citenamefont{Branco, Ferreira,
  Lavoura, Rebelo, Sher et~al.}}]{Branco:2012hc}
\bibinfo{author}{\bibfnamefont{G.}~\bibnamefont{Branco}},
  \bibinfo{author}{\bibfnamefont{P.}~\bibnamefont{Ferreira}},
  \bibinfo{author}{\bibfnamefont{L.}~\bibnamefont{Lavoura}},
  \bibinfo{author}{\bibfnamefont{M.}~\bibnamefont{Rebelo}},
  \bibinfo{author}{\bibfnamefont{M.}~\bibnamefont{Sher}}, \bibnamefont{et~al.},
  \bibinfo{journal}{Phys.Rept.} \textbf{\bibinfo{volume}{516}},
  \bibinfo{pages}{1} (\bibinfo{year}{2012}), \eprint{1106.0034}.

\bibitem[{\citenamefont{Ecker et~al.}(1987)\citenamefont{Ecker, Grimus, and
  Neufeld}}]{Ecker:1987qp}
\bibinfo{author}{\bibfnamefont{G.}~\bibnamefont{Ecker}},
  \bibinfo{author}{\bibfnamefont{W.}~\bibnamefont{Grimus}}, \bibnamefont{and}
  \bibinfo{author}{\bibfnamefont{H.}~\bibnamefont{Neufeld}},
  \bibinfo{journal}{J.Phys.} \textbf{\bibinfo{volume}{A20}},
  \bibinfo{pages}{L807} (\bibinfo{year}{1987}).

\bibitem[{\citenamefont{Gronau et~al.}(1986)\citenamefont{Gronau, Kfir, and
  Loewy}}]{Gronau:1986xb}
\bibinfo{author}{\bibfnamefont{M.}~\bibnamefont{Gronau}},
  \bibinfo{author}{\bibfnamefont{A.}~\bibnamefont{Kfir}}, \bibnamefont{and}
  \bibinfo{author}{\bibfnamefont{R.}~\bibnamefont{Loewy}},
  \bibinfo{journal}{Phys.Rev.Lett.} \textbf{\bibinfo{volume}{56}},
  \bibinfo{pages}{1538} (\bibinfo{year}{1986}).

\bibitem[{\citenamefont{Bernabeu et~al.}(1986)\citenamefont{Bernabeu, Branco,
  and Gronau}}]{Bernabeu:1986fc}
\bibinfo{author}{\bibfnamefont{J.}~\bibnamefont{Bernabeu}},
  \bibinfo{author}{\bibfnamefont{G.}~\bibnamefont{Branco}}, \bibnamefont{and}
  \bibinfo{author}{\bibfnamefont{M.}~\bibnamefont{Gronau}},
  \bibinfo{journal}{Phys.Lett.} \textbf{\bibinfo{volume}{B169}},
  \bibinfo{pages}{243} (\bibinfo{year}{1986}).

\bibitem[{\citenamefont{Branco et~al.}(1986)\citenamefont{Branco, Lavoura, and
  Rebelo}}]{Branco:1986gr}
\bibinfo{author}{\bibfnamefont{G.~C.} \bibnamefont{Branco}},
  \bibinfo{author}{\bibfnamefont{L.}~\bibnamefont{Lavoura}}, \bibnamefont{and}
  \bibinfo{author}{\bibfnamefont{M.~N.} \bibnamefont{Rebelo}},
  \bibinfo{journal}{Phys. Lett.} \textbf{\bibinfo{volume}{B180}},
  \bibinfo{pages}{264} (\bibinfo{year}{1986}).

\bibitem[{\citenamefont{Ma and Rajasekaran}(2001)}]{Ma:2001lr}
\bibinfo{author}{\bibfnamefont{E.}~\bibnamefont{Ma}} \bibnamefont{and}
  \bibinfo{author}{\bibfnamefont{G.}~\bibnamefont{Rajasekaran}},
  \bibinfo{journal}{Phys.Rev.} \textbf{\bibinfo{volume}{D64}},
  \bibinfo{pages}{113012} (\bibinfo{year}{2001}), \eprint{hep-ph/0106291}.

\bibitem[{\citenamefont{de~Adelhart~Toorop
  et~al.}(2011)\citenamefont{de~Adelhart~Toorop, Bazzocchi, Merlo, and
  Paris}}]{Adelhart-Toorop:2010fkv3}
\bibinfo{author}{\bibfnamefont{R.}~\bibnamefont{de~Adelhart~Toorop}},
  \bibinfo{author}{\bibfnamefont{F.}~\bibnamefont{Bazzocchi}},
  \bibinfo{author}{\bibfnamefont{L.}~\bibnamefont{Merlo}}, \bibnamefont{and}
  \bibinfo{author}{\bibfnamefont{A.}~\bibnamefont{Paris}},
  \bibinfo{journal}{JHEP} \textbf{\bibinfo{volume}{1103}}, \bibinfo{pages}{035}
  (\bibinfo{year}{2011}), \bibinfo{note}{{\it The discussion of CP has been
  corrected in v4.}}, \eprint{1012.1791v3}.

\bibitem[{\citenamefont{Ferreira and Lavoura}(2011)}]{Ferreira:2011lr}
\bibinfo{author}{\bibfnamefont{P.}~\bibnamefont{Ferreira}} \bibnamefont{and}
  \bibinfo{author}{\bibfnamefont{L.}~\bibnamefont{Lavoura}}
  (\bibinfo{year}{2011}), \eprint{1111.5859}.

\bibitem[{\citenamefont{Machado et~al.}(2011)\citenamefont{Machado, Montero,
  and Pleitez}}]{Machado:2011yq}
\bibinfo{author}{\bibfnamefont{A.}~\bibnamefont{Machado}},
  \bibinfo{author}{\bibfnamefont{J.}~\bibnamefont{Montero}}, \bibnamefont{and}
  \bibinfo{author}{\bibfnamefont{V.}~\bibnamefont{Pleitez}},
  \bibinfo{journal}{Phys.Lett.} \textbf{\bibinfo{volume}{B697}},
  \bibinfo{pages}{318} (\bibinfo{year}{2011}), \eprint{1011.5855}.

\bibitem[{\citenamefont{de~Adelhart~Toorop
  et~al.}(2012)\citenamefont{de~Adelhart~Toorop, Bazzocchi, Merlo, and
  Paris}}]{Adelhart-Toorop:2010fkv4}
\bibinfo{author}{\bibfnamefont{R.}~\bibnamefont{de~Adelhart~Toorop}},
  \bibinfo{author}{\bibfnamefont{F.}~\bibnamefont{Bazzocchi}},
  \bibinfo{author}{\bibfnamefont{L.}~\bibnamefont{Merlo}}, \bibnamefont{and}
  \bibinfo{author}{\bibfnamefont{A.}~\bibnamefont{Paris}}
  (\bibinfo{year}{2012}), \eprint{1012.1791v4}.

\bibitem[{\citenamefont{Altarelli and Feruglio}(2006)}]{Altarelli:2006qy}
\bibinfo{author}{\bibfnamefont{G.}~\bibnamefont{Altarelli}} \bibnamefont{and}
  \bibinfo{author}{\bibfnamefont{F.}~\bibnamefont{Feruglio}},
  \bibinfo{journal}{Nucl.Phys.} \textbf{\bibinfo{volume}{B741}},
  \bibinfo{pages}{215} (\bibinfo{year}{2006}), \eprint{hep-ph/0512103}.

\bibitem[{\citenamefont{de~Adelhart~Toorop}(2012)}]{Adelhart-Toorop:2012PhD}
\bibinfo{author}{\bibfnamefont{R.}~\bibnamefont{de~Adelhart~Toorop}}, Ph.D.
  thesis, \bibinfo{school}{Nikhef} (\bibinfo{year}{2012}).

\bibitem[{\citenamefont{Ivanov and Vdovin}(2012)}]{Ivanov:2012fp}
\bibinfo{author}{\bibfnamefont{I.}~\bibnamefont{Ivanov}} \bibnamefont{and}
  \bibinfo{author}{\bibfnamefont{E.}~\bibnamefont{Vdovin}}
  (\bibinfo{year}{2012}), \eprint{1210.6553}.

\bibitem[{\citenamefont{Degee et~al.}(2012)\citenamefont{Degee, Ivanov, and
  Keus}}]{Degee:2012sk}
\bibinfo{author}{\bibfnamefont{A.}~\bibnamefont{Degee}},
  \bibinfo{author}{\bibfnamefont{I.~P.} \bibnamefont{Ivanov}},
  \bibnamefont{and} \bibinfo{author}{\bibfnamefont{V.}~\bibnamefont{Keus}}
  (\bibinfo{year}{2012}), \eprint{1211.4989}.

\bibitem[{\citenamefont{Frampton and Kephart}(1995)}]{Frampton:1994rk}
\bibinfo{author}{\bibfnamefont{P.}~\bibnamefont{Frampton}} \bibnamefont{and}
  \bibinfo{author}{\bibfnamefont{T.}~\bibnamefont{Kephart}},
  \bibinfo{journal}{Int.J.Mod.Phys.} \textbf{\bibinfo{volume}{A10}},
  \bibinfo{pages}{4689} (\bibinfo{year}{1995}), \eprint{hep-ph/9409330}.

\bibitem[{\citenamefont{Feruglio et~al.}(2007)\citenamefont{Feruglio, Hagedorn,
  Lin, and Merlo}}]{Feruglio:2007yq}
\bibinfo{author}{\bibfnamefont{F.}~\bibnamefont{Feruglio}},
  \bibinfo{author}{\bibfnamefont{C.}~\bibnamefont{Hagedorn}},
  \bibinfo{author}{\bibfnamefont{Y.}~\bibnamefont{Lin}}, \bibnamefont{and}
  \bibinfo{author}{\bibfnamefont{L.}~\bibnamefont{Merlo}},
  \bibinfo{journal}{Nucl.Phys.} \textbf{\bibinfo{volume}{B775}},
  \bibinfo{pages}{120} (\bibinfo{year}{2007}), \eprint{hep-ph/0702194}.

\bibitem[{\citenamefont{Antusch et~al.}(2011)\citenamefont{Antusch, King, Luhn,
  and Spinrath}}]{Antusch:2011sx}
\bibinfo{author}{\bibfnamefont{S.}~\bibnamefont{Antusch}},
  \bibinfo{author}{\bibfnamefont{S.~F.} \bibnamefont{King}},
  \bibinfo{author}{\bibfnamefont{C.}~\bibnamefont{Luhn}}, \bibnamefont{and}
  \bibinfo{author}{\bibfnamefont{M.}~\bibnamefont{Spinrath}},
  \bibinfo{journal}{Nucl.Phys.} \textbf{\bibinfo{volume}{B850}},
  \bibinfo{pages}{477} (\bibinfo{year}{2011}), \eprint{1103.5930}.

\bibitem[{\citenamefont{de~Medeiros~Varzielas
  et~al.}(2007{\natexlab{a}})\citenamefont{de~Medeiros~Varzielas, King, and
  Ross}}]{de-Medeiros-Varzielas:2007gf}
\bibinfo{author}{\bibfnamefont{I.}~\bibnamefont{de~Medeiros~Varzielas}},
  \bibinfo{author}{\bibfnamefont{S.}~\bibnamefont{King}}, \bibnamefont{and}
  \bibinfo{author}{\bibfnamefont{G.}~\bibnamefont{Ross}},
  \bibinfo{journal}{Phys.Lett.} \textbf{\bibinfo{volume}{B648}},
  \bibinfo{pages}{201} (\bibinfo{year}{2007}{\natexlab{a}}),
  \eprint{hep-ph/0607045}.

\bibitem[{\citenamefont{Parattu and Wingerter}(2011)}]{Parattu:2010cy}
\bibinfo{author}{\bibfnamefont{K.~M.} \bibnamefont{Parattu}} \bibnamefont{and}
  \bibinfo{author}{\bibfnamefont{A.}~\bibnamefont{Wingerter}},
  \bibinfo{journal}{Phys.Rev.} \textbf{\bibinfo{volume}{D84}},
  \bibinfo{pages}{013011} (\bibinfo{year}{2011}), \eprint{1012.2842}.

\bibitem[{\citenamefont{Holthausen and Schmidt}(2012)}]{Holthausen:2011vd}
\bibinfo{author}{\bibfnamefont{M.}~\bibnamefont{Holthausen}} \bibnamefont{and}
  \bibinfo{author}{\bibfnamefont{M.~A.} \bibnamefont{Schmidt}},
  \bibinfo{journal}{JHEP} \textbf{\bibinfo{volume}{1201}}, \bibinfo{pages}{126}
  (\bibinfo{year}{2012}), \eprint{1111.1730}.

\bibitem[{\citenamefont{Holthausen et~al.}(2012)\citenamefont{Holthausen,
  Lindner, and Schmidt}}]{Holthausen:2012wz}
\bibinfo{author}{\bibfnamefont{M.}~\bibnamefont{Holthausen}},
  \bibinfo{author}{\bibfnamefont{M.}~\bibnamefont{Lindner}}, \bibnamefont{and}
  \bibinfo{author}{\bibfnamefont{M.~A.} \bibnamefont{Schmidt}}
  (\bibinfo{year}{2012}), \eprint{1211.5143}.

\bibitem[{\citenamefont{Pakvasa and Sugawara}(1978)}]{Pakvasa:1977in}
\bibinfo{author}{\bibfnamefont{S.}~\bibnamefont{Pakvasa}} \bibnamefont{and}
  \bibinfo{author}{\bibfnamefont{H.}~\bibnamefont{Sugawara}},
  \bibinfo{journal}{Phys.Lett.} \textbf{\bibinfo{volume}{B73}},
  \bibinfo{pages}{61} (\bibinfo{year}{1978}).

\bibitem[{\citenamefont{Pakvasa and Sugawara}(1979)}]{Pakvasa:1978tx}
\bibinfo{author}{\bibfnamefont{S.}~\bibnamefont{Pakvasa}} \bibnamefont{and}
  \bibinfo{author}{\bibfnamefont{H.}~\bibnamefont{Sugawara}},
  \bibinfo{journal}{Phys.Lett.} \textbf{\bibinfo{volume}{B82}},
  \bibinfo{pages}{105} (\bibinfo{year}{1979}).

\bibitem[{\citenamefont{Adler}(1999)}]{Adler:1998as}
\bibinfo{author}{\bibfnamefont{S.~L.} \bibnamefont{Adler}},
  \bibinfo{journal}{Phys.Rev.} \textbf{\bibinfo{volume}{D59}},
  \bibinfo{pages}{015012} (\bibinfo{year}{1999}), \eprint{hep-ph/9806518}.

\bibitem[{\citenamefont{Luhn et~al.}(2007{\natexlab{a}})\citenamefont{Luhn,
  Nasri, and Ramond}}]{Luhn:2007sy}
\bibinfo{author}{\bibfnamefont{C.}~\bibnamefont{Luhn}},
  \bibinfo{author}{\bibfnamefont{S.}~\bibnamefont{Nasri}}, \bibnamefont{and}
  \bibinfo{author}{\bibfnamefont{P.}~\bibnamefont{Ramond}},
  \bibinfo{journal}{Phys.Lett.} \textbf{\bibinfo{volume}{B652}},
  \bibinfo{pages}{27} (\bibinfo{year}{2007}{\natexlab{a}}), \eprint{0706.2341}.

\bibitem[{\citenamefont{Hagedorn et~al.}(2009)\citenamefont{Hagedorn, Schmidt,
  and Smirnov}}]{Hagedorn:2008bc}
\bibinfo{author}{\bibfnamefont{C.}~\bibnamefont{Hagedorn}},
  \bibinfo{author}{\bibfnamefont{M.~A.} \bibnamefont{Schmidt}},
  \bibnamefont{and} \bibinfo{author}{\bibfnamefont{A.~Y.}
  \bibnamefont{Smirnov}}, \bibinfo{journal}{Phys.Rev.}
  \textbf{\bibinfo{volume}{D79}}, \bibinfo{pages}{036002}
  (\bibinfo{year}{2009}), \eprint{0811.2955}.

\bibitem[{\citenamefont{de~Medeiros~Varzielas
  et~al.}(2007{\natexlab{b}})\citenamefont{de~Medeiros~Varzielas, King, and
  Ross}}]{deMedeirosVarzielas:2005qg}
\bibinfo{author}{\bibfnamefont{I.}~\bibnamefont{de~Medeiros~Varzielas}},
  \bibinfo{author}{\bibfnamefont{S.}~\bibnamefont{King}}, \bibnamefont{and}
  \bibinfo{author}{\bibfnamefont{G.}~\bibnamefont{Ross}},
  \bibinfo{journal}{Phys.Lett.} \textbf{\bibinfo{volume}{B644}},
  \bibinfo{pages}{153} (\bibinfo{year}{2007}{\natexlab{b}}),
  \eprint{hep-ph/0512313}.

\bibitem[{\citenamefont{Luhn et~al.}(2007{\natexlab{b}})\citenamefont{Luhn,
  Nasri, and Ramond}}]{Luhn:2007ul}
\bibinfo{author}{\bibfnamefont{C.}~\bibnamefont{Luhn}},
  \bibinfo{author}{\bibfnamefont{S.}~\bibnamefont{Nasri}}, \bibnamefont{and}
  \bibinfo{author}{\bibfnamefont{P.}~\bibnamefont{Ramond}},
  \bibinfo{journal}{J.Math.Phys.} \textbf{\bibinfo{volume}{48}},
  \bibinfo{pages}{073501} (\bibinfo{year}{2007}{\natexlab{b}}),
  \eprint{hep-th/0701188}.

\bibitem[{\citenamefont{Escobar and Luhn}(2009)}]{Escobar:2008vc}
\bibinfo{author}{\bibfnamefont{J.}~\bibnamefont{Escobar}} \bibnamefont{and}
  \bibinfo{author}{\bibfnamefont{C.}~\bibnamefont{Luhn}},
  \bibinfo{journal}{J.Math.Phys.} \textbf{\bibinfo{volume}{50}},
  \bibinfo{pages}{013524} (\bibinfo{year}{2009}), \eprint{0809.0639}.

\end{thebibliography}

\end{document}